\documentclass[reqno]{amsart}

\usepackage{amsthm,amsmath,amsfonts,amssymb,epsfig,graphicx,latexsym,psfrag,float}

\newtheorem{definition}{Definition}

\newtheorem{remark}[definition]{Remark}

\newtheorem{notation}[definition]{Notation}

\renewcommand{\theequation}{\arabic{section}.\arabic{equation}}

\newcommand{\Real}{\mbox{$\mathbb{R}$}}

\DeclareSymbolFont{msbm}{U}{msb}{m}{n}
\DeclareMathSymbol{\R}{\mathalpha}{msbm}{'122}

\begin{document}

\title[Finite volume approach for instationary viscous Cosserat rods]
{Finite volume approach for the instationary Cosserat rod model describing the spinning of viscous jets}

\author{Walter Arne}
%\address{W.~Arne, Universit\"at Kassel, FB Mathematik und Naturwissenschaften, Heinrich Plett Str.\ 40, D-34132 Kassel; Fraunhofer ITWM, Fraunhofer Platz 1, D-67663 Kaiserslautern, Germany}
%\email{arne@itwm.fhg.de}
\author{Nicole Marheineke}
%\address{N.~Marheineke, FAU Erlangen-N\"urnberg, Lehrstuhl Angewandte Mathematik I, Cauerstr.\ 11, D-91058 Erlangen, Germany}
%\email{marheineke@am.uni-erlangen.de}
\author{Andreas Meister}
%\address{A.~Meister, Universit\"at Kassel, FB Mathematik und Naturwissenschaften, Heinrich Plett Str.\ 40, D-34132 Kassel Kassel, Germany}
%\email{meister@mathematik.uni-kassel.de}
\author{Raimund Wegener} 
%\address{R.~Wegener, Fraunhofer ITWM, Fraunhofer Platz 1, D-67663 Kaiserslautern, Germany}
%\email{wegener@itwm.fhg.de}

%% Footnote for date and authors affiliations (to appear on title page)
\date{\today\\
W.~Arne, Universit\"at Kassel, FB Mathematik und Naturwissenschaften, Heinrich Plett Str.\ 40, D-34132 Kassel; 
Fraunhofer ITWM, Fraunhofer Platz 1, D-67663 Kaiserslautern, Germany. \texttt{arne@itwm.fhg.de}\\
N.~Marheineke, FAU Erlangen-N\"urnberg, Lehrstuhl Angewandte Mathematik I, Cauerstr.\ 11, D-91058 Erlangen, Germany. 
\texttt{marheineke@am.uni-erlangen.de} (\textit{corresponding author})\\
A.~Meister, Universit\"at Kassel, FB Mathematik und Naturwissenschaften, Heinrich Plett Str.\ 40, D-34132 Kassel, Germany.
\texttt{meister@mathematik.uni-kassel.de}\\
R.~Wegener, Fraunhofer ITWM, Fraunhofer Platz 1, D-67663 Kaiserslautern, Germany.
\texttt{wegener@itwm.fhg.de}}

\begin{abstract} 
The spinning of slender viscous jets can be described asymptotically by one-dimen\-sional models that consist of systems of partial and ordinary differential equations. Whereas the well-established string models possess only solutions for certain choices of parameters and set-ups, the more sophisticated rod model that can be considered as $\epsilon$-regularized string is generally applicable. But containing the slenderness ratio $\epsilon$ explicitely in the equations complicates the numerical treatment. In this paper we present the first instationary simulations of a rod in a rotational spinning process for arbitrary parameter ranges with free and fixed jet end, for which the hitherto investigations longed. So we close an existing gap in literature. The numerics is based on a finite volume approach with mixed central, up- and down-winded differences, the time integration is performed by stiff accurate Radau methods.
\end{abstract}

\maketitle

\keywords{{\sc Keywords.} Rotational spinning process, viscous fiber, special Cosserat theory, partial differential algebraic equations, quaternions, finite volume scheme}

%\subjclass{{\sc AMS-Classification.} 65Mxx, 76-xx, 35Q35}

%%%%%%%%%%%%%%%%%%%%%%%%%%%%%%%%%%%%%%%%%%%%%%%%%%%%%%%%%%%%%%%%%%%%%%%%%%%%%%%%%%%%%%%
\section{Introduction}

The understanding of jet spinning is of interest in many industrial applications, including for example drawing, tapering and spinning of glass and polymer fibers \cite{pearson:b:1985,klar:p:2009,forest:p:2001} and pellet manufacturing \cite{decent:p:2009,parau:p:2007}. Considering the spinning of highly viscous fluids, the unrestricted motion of an instationary jet's center-line is an important feature, as experiments show (see "break-up mode 4" by Wong et al.\ \cite{wong:p:2004}). In the context of slender-body theory there exist two classes of one-dimensional models for the numerical simulation of such a jet, string and rod models, \cite{antman:b:2006, buckmaster:p:1975, entov:p:1984, yarin:b:1993}. The string models are asymptotic systems of leading order that result from the three-dimensional free boundary value problems of Newtonian fluid flows in a strict systematic derivation using expansions in the slenderness ratio $\epsilon$ ($\epsilon \ll 1$). They consist of balance laws for mass and linear momentum. The more sophisticated rod models also possess an angular momentum balance. The rod models follow from the cross-sectional averaging of the underlying three-dimensional balance equations, assuming that the displacement field in each cross-section can be expressed in terms of a finite number of vector- and tensor-valued quantities. The constitutive elements of a (special) Cosserat rod are a curve and a director triad specifying the position (center-line) and the orientation of the cross-sections, respectively. The one-dimensional material and geometrical laws that are needed to close the model are heuristically motivated. The Cosserat rod model is no asymptotic system of leading order but contains the slenderness ratio $\epsilon$ explicitely in the angular momentum balance. As the rod reduces to a string as $\epsilon \rightarrow 0$, the Cosserat rod can be considered as $\epsilon$-regularized string. This regularization allows the rod to overcome limitations that the strings have in their applicability, in particular when dealing with time-dependencies. In this paper we present the first instationary simulations of a Cosserat rod in a rotational spinning process for arbitrary parameter ranges with free and fixed jet end, where the string models failed so far. 

A string model for the jet dynamics was recently deduced in a rigorous slender-body asymptotics from the three-dimensional free boundary value problem given by the incompressible Navier-Stokes equations, \cite{marheineke:p:2009}. Accounting for inner viscous transport, surface tension and placing no restrictions on either the motion or the shape of the jet's center-line, it generalizes the previously developed string models for straight \cite{cummings:p:1999,dewynne:p:1994,dewynne:p:1992} and curved \cite{decent:p:2002,panda:p:2008,wallwork:p:2002} center-lines (for a detailed survey of literature see \cite{marheineke:p:2009}). The numerical results investigating the effects of viscosity, surface tension, inertia and gravity on the jet behavior coincide well with the experiments of \cite{wong:p:2004}. However, the applicability of the string model turned out to be restricted to certain parameter ranges. Neglecting surface tension and gravity, already for jets in a stationary, rotational two-dimensional scenario no "physically relevant" solutions exist for ${\rm Re}{\rm Rb}^2<1$ with Reynolds number ${\rm Re}^{-1}\ll 1$ and Rossby number ${\rm Rb} \ll 1$ according to \cite{goetz:p:2008,arne:p:2010}. The numerical evidence of this inviscid bound was specified analytically in \cite{arne:p:2011}; it is ${\rm Re}{\rm Rb}^2=3/(2\min_i|\lambda_i|^3) \approx 1.4$ with $\lambda_i$ root of the Airy Prime function. The restricted applicability / validity results from a non-removable singularity in the model equations due to an inconsistency entering with the asymptotically deduced boundary conditions that prescribe the jet tangent at the spinning nozzle. This limitation can be overcome by a modification of the closure conditions; the boundary condition is omitted in favor of an interface condition that avoids the occurrence of the singularity and ensures the regularity of the string quantities. This change implies a different string model describing an other jet regime. For gravitational spinning Hlod et al.\ \cite{hlod:p:2007,hlod:p:2012} distinguish between three compatible disjoint jet regimes, i.e.\ inertial, viscous-inertial and viscous regimes, that they successfully investigated using the string equations with appropriately chosen closure conditions. The classification of the regimes is transferable to rotational spinning, \cite{arne:p:2011}. But, here the regimes do not cover the whole parameter range. Already for the stationary, rotational two-dimensional scenario an existence gap of the string solutions is observed for ${\rm Re} \ll1$, ${\rm Rb}\ll 1$,  \cite{arne:p:2011}. It is handed over to the instationary simulations that break down for viscous fiber jets under very high rotations as they occur in industrial production processes of glass wool \cite{marheineke:p:2009}. When surface tension, aerodynamic forces and temperature-dependent viscosity are included, the question of existence and solvability becomes much more difficult or even impossible to answer. In non-stationary spinning processes the jet behavior and regime might also change over time. To handle this difficulty Hlod \cite{hlod:d:2009} investigated a numerical (ad hoc) switching of the closure conditions in the simulations. The heuristic approach is motivated by the embedding of the instationary string equations into the hyperbolic theory of characteristics under certain assumptions. However, the studies remain dissatisfactorily in view of real applications.
 
The viscous Cosserat rod theory raises hope to open the parameter ranges of practical interest and time-dependencies to simulation and optimization.
For the coiling of a viscous jet falling onto a rigid substrate Ribe \cite{ribe:p:2004,ribe:p:2006a} proposed a rod model with dynamic center-line that allows for stretching, bending and twisting and that is clearly superior to the strings in the application of a fluid-mechanical "sewing machine", see stationary simulations in \cite{ribe:p:2006b,chiu:p:2006} and stability analysis in \cite{ribe:p:2006,morris:p:2008}. Based on these studies and embedded in the special Cosserat theory, we developed a modified incompressible rod model for spinning \cite{arne:p:2010} that reduces asymptotically to the string equations of \cite{marheineke:p:2009} for a vanishing slenderness parameter $\epsilon$. It not only covers the string models, but also overcomes all thitherto restrictions. In case of stationarity the rod solutions exist for all parameter ranges and spinning scenarios without any exceptions, and the existing string solutions belonging to the different jet regimes (different closure conditions) are their asymptotic limits as $\epsilon \rightarrow 0$, see convergence results in \cite{arne:p:2011}. Corresponding stationary rod simulations have been successfully applied in the study and design of glass wool production processes, \cite{arne:p:2011a,marheineke:p:2012}. The instationary rod is described by a system of partial and ordinary differential equations that becomes stiff for small $\epsilon$ and hence requires a careful numerical treatment. Apart from this structure a further numerical challenge lies in the accurate realization of the angular momentum effects which involves the conservation of the orthonormal director triad that is attached to the jet`s center-line and characterizes the orientation of the cross-sections over time. Posing, in favor of a material law for the inner forces, a modified Kirchhoff constraint $\boldsymbol{\tau}=e \mathbf{d_3}$ that relates the jet tangent $\boldsymbol{\tau}$ and the director $\mathbf{d_3}$ via the elongation $e$, the vector-valued angular velocity can be expressed in terms of the tangent and the scalar-valued spin (tangential angular speed). So the angular momentum balance becomes scalar-valued and the temporal evolution of the triad redundant, as the other components can be computed a posteriori. Motivated from the numerics of elastic Kirchhoff beams Audoly et al.\ \cite{audoly:p:2012} just recently developed a discrete geometric Lagrangian approach and performed instationary simulations for a jet lay-down (see also \cite{brun:p:2012}). Thereby, they studied the effect of inertia in the angular momentum balance. Its neglect simplifies the numerics due to a change of the equations' structure. The numerical handling of a free jet end was addressed as open question and topic of future research. 

In this paper, we propose a finite volume approach with mixed central, up- and down-winded differences for the instationary viscous rod with free and fixed end in Lagrangian and Eulerian parameterization, respectively. The time integration is performed by stiff accurate Radau methods taking into account the differential-algebraic character of the system. The rotational tensor associated to the orthonormal director triad is realized using unit quaternions. The approach enables the simulation of two-dimensional and three-dimensional rotational spinning for arbitrary (unrestricted) parameter ranges for which the hitherto investigations longed and failed. We deal with inflow-outflow set-ups with fixed domain and inflow set-ups with time-dependent enlarging domain and discuss the results in comparison to stationary rod \cite{arne:p:2010,arne:p:2011} and instationary string simulations \cite{marheineke:p:2009,panda:d:2006,panda:p:2008}, respectively. So this paper closes a gap in existing literature.

The paper is structured as follows. After a short survey of the special Cosserat theory for viscous jets in Section~\ref{sec:2}, we formulate the instationary rod model in Lagrangian and Eulerian parameterizations.  For the resulting initial-boundary value problems we develop a finite volume approach with Radau time integration in Section~\ref{sec:3}. In Section~\ref{sec:4} we perform numerical simulations for the two practically relevant spinning set-ups of enlarging and fixed flow domains and investigate the instationary effects. By allowing for the study of all parameter ranges, the rod model shows its large potential in view of simulating and optimizing non-stationary three-dimensional rotational spinning processes in industrial applications in future.

%%%%%%%%%%%%%%%%%%%%%%%%%%%%%%%%%%%%%%%%%%%%%%%%%%%%%%%%%%
  
\section{Special Cosserat theory for viscous jets}\label{sec:2}

A jet is a slender long body. Because of its geometry with slenderness ratio $\epsilon$ ($\epsilon\ll1$), its dynamics might be reduced to a one-dimensional description by averaging the underlying balance laws over its cross-sections. The procedure is based on the assumption that the displacement field in each cross-section can be expressed in terms of a finite number of vector- and tensor-valued quantities. The special Cosserat rod theory consists hereby of only two constitutive elements, a curve specifying the position and an orthonormal director triad characterizing the orientation of the cross-sections, for details see \cite{antman:b:2006}. It represents a general framework that might be applicable to all materials and set-ups. The core of the description are physically reasonable one-dimensional geometrical and material laws. In this work we use the incompressible viscous rod model derived in \cite{arne:p:2010}, whose asymptotic limit as $\epsilon\rightarrow 0$ are the string equations of \cite{marheineke:p:2009,panda:p:2008}. Since the model equations can be formulated in various ways depending on the choice of parameterization/coordinates (Lagrangian or Eulerian), basis (invariant, director or outer basis), reference system (fixed or rotational), set-up (time-dependent or -independent flow domain, acting forces, 2d or 3d), dimensions (with dimensions or dimensionless) and so on, we start our introduction with the general invariant description of the rod in a Lagrangian parameterization from which all other re-formulations can be straightforward computed. In addition, we explicitly state the model formulations that are relevant in the considered spinning application and that form the basis for the development of our numerical approach, i.e.\ inflow set-up with enlarging domain (free jet end) in Lagrangian parameterization as well as inflow-outflow set-up with fixed domain in Eulerian parameterization. This choice of parameterization yields initial-boundary value problems on given computational domains in both cases, which facilitates the numerical treatment.

%%%%%%%%%%%
\subsection{General invariant formulation of instationary viscous Cosserat rod model}

A special Cosserat rod in the three-dimensional Euclidean space $\mathbb{E}^3$ is defined by a curve $\mathbf{r}:\mathcal{Q}\rightarrow\mathbb{E}^3$ and an orthonormal director triad $\{\mathbf{d_1},\mathbf{d_2},{\bf d_3}\}:\mathcal{Q}\rightarrow\mathbb{E}^3$ with $\mathcal{Q}=\{(\sigma,t)\in \mathbb{R}^2 \,|\, \sigma\in [\sigma_a(t),\sigma_b(t)],\, t\geq 0\}$, where $\sigma$ addresses a material cross-section (material point) of the rod. The domain of the material parameter is chosen to be time-dependent to allow for inflow/outflow boundaries and free end in the Lagrangian description. Considering the dynamics of an incompressible isothermal viscous inertial jet with circular cross-sections and constant mass density, the rod model consists of four kinematic and two dynamic equations that are equipped with specific geometrical assumptions (shape-preserving incompressibility) and material laws. Its invariant formulation reads \cite{arne:p:2010} 
\begin{align}\label{eq:L_invariant} 
\partial_t\mathbf{r} &={\bf v} \\ \nonumber
\partial_t\mathbf{d_k}&=\boldsymbol{\omega}\times\mathbf{d_k}\\\nonumber
\partial_t\boldsymbol{\tau}&=\partial_\sigma \mathbf{v}\\ \nonumber	 
\partial_t\boldsymbol{\kappa}&=\partial_\sigma \boldsymbol{\omega}+\boldsymbol{\omega}\times\boldsymbol{\kappa}\\\nonumber
\varrho A_\circ \partial_t \mathbf{v}  &= \partial_\sigma \mathbf{n} + \mathbf{k} \\ \nonumber
\varrho \partial_t\left(\mathbf{J_\circ}\cdot \frac{\boldsymbol{\omega}}{e}\right)& = \partial_\sigma \mathbf{m} + \boldsymbol{\tau}\times\mathbf{n} +\mathbf{l}
\end{align}
with
\begin{align*}
\mathbf{J_\circ}&=I_\circ(\mathbf{d_1}\otimes \mathbf{d_1}+\mathbf{d_2}\otimes \mathbf{d_2}+2\mathbf{d_3}\otimes \mathbf{d_3}), \qquad 
I_\circ=\frac{A_\circ^2}{4\pi} \\ 
\mathbf{n}\cdot \mathbf{d_3} =3\mu A_\circ \frac{\partial_\sigma \mathbf{v}}{e^2}\cdot \mathbf{d_3}, \qquad 
\mathbf{m}&= 3 \mu I_\circ \left(\mathbf{d_1}\otimes \mathbf{d_1}+\mathbf{d_2}\otimes \mathbf{d_2}+\frac{2}{3}\mathbf{d_3}\otimes \mathbf{d_3}\right) \cdot \frac{\partial_\sigma \boldsymbol{\omega}}{e^3}, \qquad 
\boldsymbol{\tau}=e\mathbf{d_3}
\end{align*}
and appropriate initial and boundary conditions. Note that system~\eqref{eq:L_invariant} includes $\partial_\sigma \mathbf{r}=\boldsymbol{\tau}$ and $\partial_\sigma \mathbf{d_k}=\boldsymbol{\kappa}\times \mathbf{d_k}$. The derivatives of the curve $\mathbf{r}$ with respect to time and material parameter are the velocity $\mathbf{v}$ and the tangent field $\boldsymbol{\tau}$. Due to the orthonormality of the directors $\{\mathbf{d_1},\mathbf{d_2},{\bf d_3}\}$, their derivatives imply the existence of the angular velocity $\boldsymbol{\omega}$ and the generalized curvature $\boldsymbol{\kappa}$. Assuming sufficient regularity, 
${\bf v}, \boldsymbol{\tau}$ as well as $\boldsymbol{\omega}, \boldsymbol{\kappa}$ are related according to the stated compatibility conditions (third and fourth equations in \eqref{eq:L_invariant}). The dynamic equations are the balance equations for linear and angular momentum with external loads $\mathbf{k}$, ${\bf l}$ (body force and body couple line density) coming from the considered application. In case of temperature dependencies a corresponding balance can be added straightforward, cf.\ \cite{arne:p:2011a}. The curve $\mathbf{r}$ is here chosen as the mass-associated center-line. The line density $\varrho A_\circ$ with constant mass density $\varrho$ as well as the polar moment of inertia $I_\circ$ refer to the referential circular cross-sectional area $A_\circ$ and are hence time-independent. The incompressibility leads to a shrinking of the cross-sections when stretching the body. During the deformation their shapes are assumed to be retained. This is incorporated in the geometrical model for the angular momentum being linear in $\boldsymbol{\omega}$ with dilatation measure $e>0$.
In consequence the actual cross-sectional area and moment of inertia are given by $A=A_\circ/e$ and $I=I_\circ/e^2$. The reference area $A_\circ$ could be replaced by  $A$ in \eqref{eq:L_invariant}, which requires the adding of the evolution equation $\partial_t (eA)=0$.
The algebraic relation $\boldsymbol{\tau}=e\mathbf{d_3}$ that determines the tangent via $e$ represents a modified Kirchhoff constraint allowing for extensibility. By its introduction, the normal contact force components $\mathbf{n}\cdot\mathbf{d_1}$, $\mathbf{n}\cdot\mathbf{d_2}$ become Lagrangian multipliers (variables of the system). The tangential contact force $\mathbf{n}\cdot\mathbf{d_3}$ and the contact couple $\mathbf{m}$ are specified by linear material laws in the spatial derivatives of the linear and angular velocities (strain rates) with dynamic jet viscosity $\mu$, \cite{ribe:p:2004,ribe:p:2006a}. Note that for the discussion and a better understanding of the geometrical assumptions and material laws, it is most convenient to formulate the rod model \eqref{eq:L_invariant} in the director basis $\{\mathbf{d_1},\mathbf{d_2},\mathbf{d_3}\}$ as we will do later on. Summing up, the variables of the rod model \eqref{eq:L_invariant} are $\mathbf{r}$, $\{\mathbf{d_1}, \mathbf{d_2}, \mathbf{d_3}\}$, $e$, $\boldsymbol{\kappa}$, $\mathbf{v}$, $\boldsymbol{\omega}$, $\mathbf{n}\cdot \mathbf{d_1}$ and $\mathbf{n}\cdot \mathbf{d_2}$.

\begin{remark}[Impact of modified Kirchhoff constraint]\label{rem:Kirchhoff}
The applied Kirchhoff constraint $\boldsymbol{\tau}=e\mathbf{d_3}$ poses a geometric relation between curve and director triad in favor of a material law for the force components $\mathbf{n}\cdot \mathbf{d_1}$, $\mathbf{n}\cdot \mathbf{d_2}$. It allows the reduction of the unknown vector-valued angular velocity $\boldsymbol{\omega}$ to the scalar spin $W=\boldsymbol{\omega}\cdot\mathbf{d_3}$, i.e.\ 
\begin{align*}                                                                                                                                                                                                                                                                                                                  \boldsymbol{\omega}=\frac{\boldsymbol{\tau}}{e}\times \partial_t \left(\frac{\boldsymbol{\tau}}{e}\right)+W\frac{\boldsymbol{\tau}}{e}.                                                                                                                                                                                                                                                                                                                \end{align*}
A respective reformulation of \eqref{eq:L_invariant} involves a scalar-valued angular momentum balance and renounces the evaluation of the director triad, but it changes the clear system structure towards mixed derivatives. For this centerline-spin representation Audoly et al.\ \cite{audoly:p:2012} proposed a discrete geometric Lagrangian method (that is inspired by \cite{langer:p:1996} for the analogous centerline-angle representation of an elastic Kirchhoff beam), handling the derivatives algorithmically. Thereby, they neglected the influence of inertia in the angular momentum balance, cf.\ Remark~\ref{rem:simpler_model}. We stand for system \eqref{eq:L_invariant}, since its composition of partial and ordinary differential equations is well suited for (standard) finite volume schemes and stiff accurate Runge-Kutta methods whose convergence results and performance are well-known, see Section~\ref{sec:3}. 
\end{remark}

\begin{remark}[Model simplifications]\label{rem:simpler_model}
In case of negligible inertia $\varrho \partial_t(\mathbf{J_\circ}\cdot {\boldsymbol{\omega}}/{e})=\mathbf{0}$ and no outer couple $\mathbf{l}=\mathbf{0}$ in the angular momentum balance, the viscous rod system \eqref{eq:L_invariant} with modified Kirchhoff constraint reduces to
\begin{align*}
\partial_t \mathbf{r}&=\mathbf{v}\\
\partial_t \boldsymbol{\tau}&=\partial_\sigma \mathbf{v}\\
\varrho A_\circ \partial_t \mathbf{v}&=\partial_\sigma \mathbf{n}+\mathbf{k}\\
\partial_\sigma W&=\frac{e^3}{2\mu I_\circ}M-\frac{1}{e^2}\partial_\sigma \mathbf{v}\cdot (\boldsymbol{\tau}\times \partial_\sigma \boldsymbol{\tau})\\
\partial_\sigma M &= \frac{3\mu I_\circ}{e^5}\left(\frac{W}{e^2}\|\boldsymbol{\tau}\times\partial_\sigma \boldsymbol{\tau}\|^2-\partial_\sigma\frac{(\partial_\sigma \mathbf{v})^\perp}{e}\cdot(\boldsymbol{\tau}\times \partial_\sigma \boldsymbol{\tau})\right),
\end{align*}
treating the scalar tangential angular speed / spin $W$ (cf.\ Remark~\ref{rem:Kirchhoff}) and tangential contact couple component $M$ as variables. The respective contact force $\mathbf{n}$ becomes with its tangential component $N=3\mu A_\circ \partial_\sigma \mathbf{v}\cdot \boldsymbol{\tau}/e^3$ 
\begin{align*}
\mathbf{n}&=\frac{N}{e}\boldsymbol{\tau} 
-\frac{1}{e} \partial_\sigma \left(\frac{3\mu I_\circ}{e^3}\left(\partial_\sigma \frac{(\partial_\sigma \mathbf{v})^\perp}{e}
+\frac{1}{e^3}((\partial_\sigma \mathbf{v})^\perp\cdot \partial_\sigma \boldsymbol{\tau})\boldsymbol{\tau}
-\frac{W}{e^2} \boldsymbol{\tau}\times \partial_\sigma \boldsymbol{\tau}\right)\right) \\
&\quad +\frac{M}{e^3} \boldsymbol{\tau}\times \partial_\sigma \boldsymbol{\tau}
- \frac{3\mu I_\circ}{e^6}\left((\partial_\sigma \frac{(\partial_\sigma \mathbf{v})^\perp}{e}\cdot \partial_\sigma \boldsymbol{\tau})\boldsymbol{\tau} 
+\frac{1}{e^3}((\partial_\sigma \mathbf{v})^\perp\cdot \partial_\sigma \boldsymbol{\tau})(\boldsymbol{\tau}\cdot \partial_\sigma \boldsymbol{\tau})\boldsymbol{\tau}\right)\\
e&=\|\boldsymbol{\tau}\|.
\end{align*}
Here, $\mathbf{z}^\perp=\mathbf{z}-(\mathbf{z}\cdot \boldsymbol{\tau})\boldsymbol{\tau}/\|\boldsymbol{\tau}\|^2$ for any arbitrary vector $\mathbf{z}\in \mathbb{E}^3$.

The string system that is the asymptotic slenderness limit of the rod \cite{arne:p:2011} has the form 
\begin{align*}
\partial_t \mathbf{r}&=\mathbf{v}\\
\partial_t \boldsymbol{\tau}&=\partial_\sigma \mathbf{v}\\
\varrho A_\circ \partial_t \mathbf{v}&=\partial_\sigma\left(\frac{N}{\|\boldsymbol{\tau}\|}\boldsymbol{\tau}\right)+\mathbf{k}\,.
\end{align*}
with $N$ given as above.
\end{remark}

System \eqref{eq:L_invariant} is written in a Lagrangian setting. Thereby, the material parameterization might be determined up to orientation and a constant by using an arc-length parameterized reference confi\-guration. Alternatively, any other time-dependent parameterization can be used for the formulation of the model, defined via an orientated bijective mapping
\begin{align*}
S(\cdot,t):[\sigma_a(t),\sigma_b(t)]\rightarrow[S(\sigma_a(t),t),S(\sigma_b(t),t)]=[s_a(t),s_b(t)], \quad \sigma \mapsto S(\sigma,t).
\end{align*}
Assuming sufficient regularity, a scalar convective velocity $u$ and a spatial Jacobian $j$ belong to $S$:
\begin{align*}
\partial_t S(\sigma,t)=u(S(\sigma,t),t), \quad  \partial_\sigma S(\sigma,t)=j(\sigma,t)>0, \,\,\, \text{ with } \,\,\,
\partial_s u(S(\sigma,t),t)=\frac{\partial_t j}{j}(\sigma,t).
\end{align*}
The re-parameterization of all fields carries convective terms with speed $u$ into \eqref{eq:L_invariant}. Choosing $u=0$ implies the material description. Instead of imposing $u$ explicitly, also a constraint can be prescribed such that $u$ becomes the associated Lagrangian multiplier and hence an additional unknown of the system. The mostly used constraint is the arc-length parameterization of the jet curve for all times, yielding an Eulerian setting. Here, $e=j$ coincide due to the Kirchhoff constraint. Moreover, $\partial_t S(\sigma,t)= u(S(\sigma,t),t)$ prescribes the rate of change of the arc-length $S(\sigma,t)$ to the material point $\sigma$; $e$ is a measure for the strain and $\partial_s u(S(\sigma,t),t)=(\partial_t e/e)(\sigma,t)$ the corresponding relative strain rate. The Eulerian (spatial) description is certainly the most intuitive one for flow problems and allows for the transition to stationary considerations.

%%%%%%%%%%%%
\subsection{Rotational spinning process -- two relevant set-ups}

\begin{figure}[tb]
\psfrag{noz}{nozzle}
\psfrag{drum}{drum}
\psfrag{jet}{\hspace*{0.4cm}jet}
\psfrag{om}{$\Omega$}
\psfrag{g}{${\bf k}$}
\psfrag{a1}{\hspace*{-0.2cm}${\bf a_1}={\bf d_1}$}
\psfrag{a2}{${\bf a_2}$}
\psfrag{a3}{${\bf a_3}$}
\psfrag{a}{$\alpha$}
\psfrag{r0}{\hspace*{-0.5cm} ${\sf \breve{r}}(0)=(R,0)$}
\psfrag{rl}{${\sf \breve{r}}(L)$}
\psfrag{R}{$R$}
\psfrag{g00}{${\bf k}={\bf 0}$}
\includegraphics[scale=0.33]{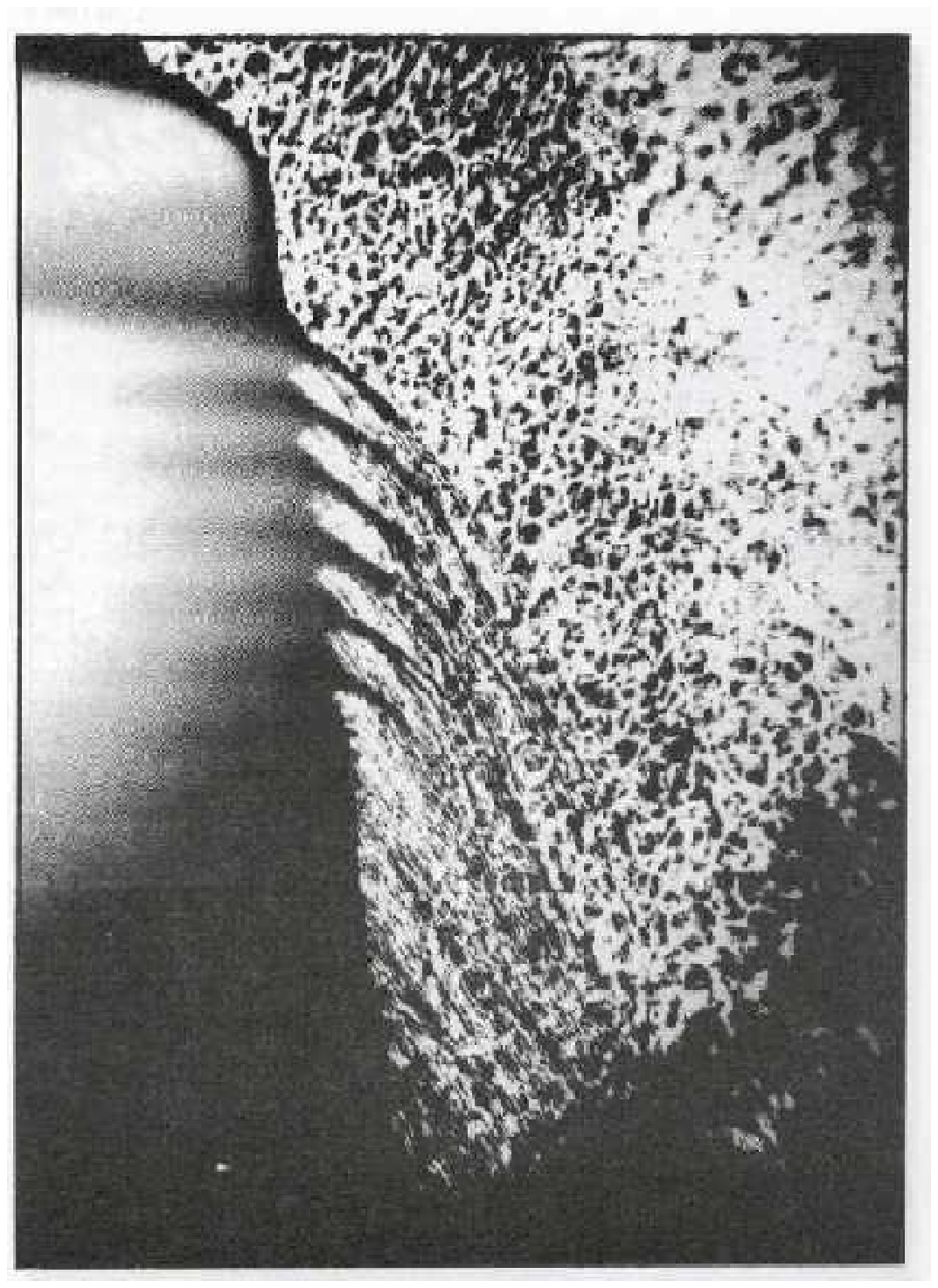}\hfill
\includegraphics[scale=0.55]{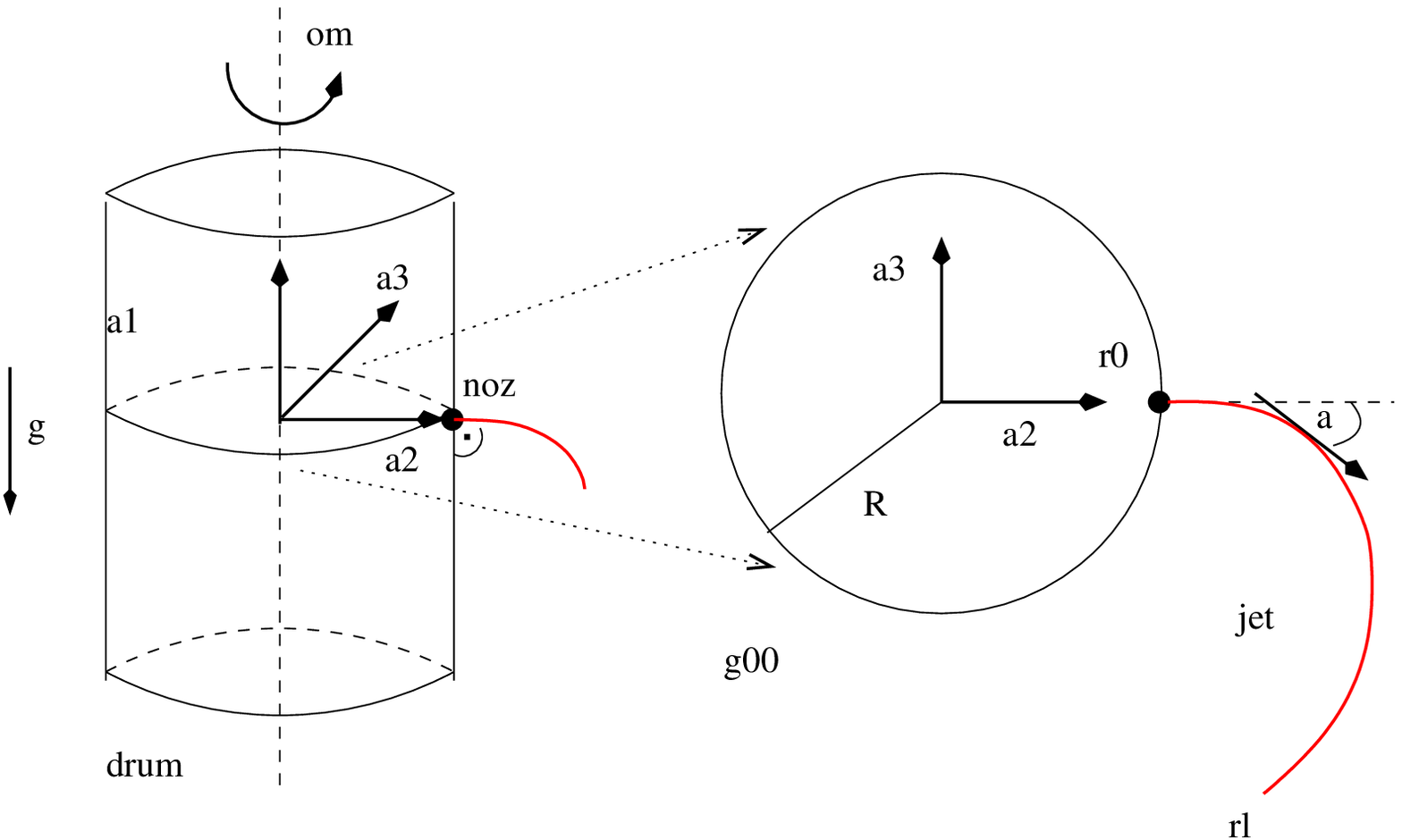}
\caption{\label{fig:1}{\it Left}: Rotational fiber spinning process, photo by industrial partner. {\it Right}: Sketch of 3d set-up and its 2d simplification under the neglect of gravity.}
\end{figure}

In rotational spinning processes  \cite{marheineke:p:2012}, viscous liquid jets leave small spinning nozzles located on the curved face of a circular cylindrical drum rotating about its symmetry axis, cf.\ Fig.~\ref{fig:1}. At the nozzle, the velocity, cross-sectional area, direction and curvature of a jet are prescribed. Starting from an initial length of zero, the extruded liquid jet grows and moves due to viscous friction, surface tension and gravity. Also aerodynamic forces might act, see e.g.\ \cite{arne:p:2011a}. In this paper we aim at a numerical treatment of the non-stationarity. For simplicity we neglect surface tension, aerodynamic forces and temperature dependencies and restrict to external loads rising from gravity, i.e.\ in the Lagrangian setting $\mathbf{k}= \varrho A_\circ g \mathbf{e_g}$ and $\mathbf{l}=\mathbf{0}$ with gravitational acceleration $g$ and direction $\mathbf{e_g}$, $\|\mathbf{e_g}\|=1$. However, note that once the numerical concept is established, the other effects can be easily added as it is already done in the stationary considerations of industrial spinning processes in \cite{marheineke:p:2012,arne:p:2011a}. 
In the following we focus on the numerical simulation of two set-ups that are important for the understanding and study of the industrial application:
\begin{itemize}
 \item[] Set-up A: \emph{inflow with enlarging domain (free jet end)}
 \item[] Set-up B: \emph{inflow-outflow with time-independent (fixed) domain}
\end{itemize}
For the inflow set-up we choose a Lagrangian (material) description, whereas the inflow-outflow set-up is formulated in an Eulerian (spatial) setting. Certainly, every set-up could also be formulated in the other parameterization, but this yields free boundary value problems. Our choice instead implies initial-boundary value problems on given computational domains, which makes the numerical treatment undeniably easier.

To describe the spinning process of interest (Fig.~\ref{fig:1}), we follow \cite{arne:p:2010} and use the reference frame that rotates with the drum. Let $\boldsymbol{\Omega}=\Omega \mathbf{e}_{\boldsymbol{\Omega}}$ with $\mathbf{e}_{\boldsymbol{\Omega}}=-\mathbf{e_g}$ be the angular frequency of the rotating device, then we introduce the rotating outer basis $\{{\bf a_1}(t),{\bf a_2}(t),{\bf a_3}(t)\}$ satisfying $\partial_t {\bf a_i}=\boldsymbol{\Omega}\times {\bf a_i}$, $i=1,2,3$. This makes the position of the nozzle and the direction of the inflow time-independent, but introduces fictitious rotational body forces and couples in the dynamic equations due to inertia. We deal with $\boldsymbol{\Omega}$-adapted velocity and angular speed, i.e.\ $\mathbf{v}_{\boldsymbol{\Omega}} = \mathbf{v}-(\boldsymbol{\Omega}\times \mathbf{r})$ and $ \boldsymbol{\omega}_{\boldsymbol{\Omega}} = \boldsymbol{\omega} -\boldsymbol{\Omega}$. Note that we skip the subscript $_{\boldsymbol{\Omega}}$ in the following to facilitate the readability. Moreover, we state the model equations in the director basis $\{\mathbf{d_1},\mathbf{d_2},\mathbf{d_3}\}$ for reasons of the material laws and geometrical assumptions, see Notation~\ref{not:1}.
The rod model for rotational spinning has eight physical parameters: jet density $\varrho$, viscosity $\mu$, length $L$, diameter $D$ and velocity $U$ at the nozzle as well as drum radius $R$, rotational frequency $\Omega$ and gravitational acceleration $g$. These induce five characteristic dimensionless numbers: Reynolds number ${\rm Re}=\varrho UR/\mu$ as ratio between inertia and viscosity, Rossby number ${\rm Rb}=U/(\Omega R)$ as ratio between inertia and rotation, Froude number $\mathrm{Fr}=U/\sqrt{gR}$ as ratio between inertia and gravity as well as $l=L/R$ and $\epsilon=D/R$ as length ratios between jet length, nozzle diameter respectively and drum radius. For the subsequent numerical investigations, we make the Lagrangian system dimensionless by scaling the quantities with the following reference values: 
\begin{align*}
\sigma_0&=r_0=R,\quad v_0=U, \quad t_0=v_0/r_0, \quad \kappa_0=1/r_0, \quad \omega_0=r_0/v_0\quad 
A_0=\pi D^2/4, \quad k_0=\varrho A_0v_0^2/r_0\\
n_0&=  \mu A_0 v_0 /r_0  = \pi \varrho v_0^2 r_0^2 \epsilon^2 / (4\mathrm{Re}),\quad 
m_0=  \mu A_0^2 v_0 /(\pi r_0^2)  = \pi \varrho v_0^2 r_0^3 \epsilon^4/(16\mathrm{Re}).
\end{align*} 
In the Eulerian setting we use alternatively $s_0=r_0$, and consistently $u_0=v_0$ for the intrinsic velocity.

\begin{notation}\label{not:1}
To an arbitrary vector field $\mathbf{z}=\sum_{i=1}^3 \breve z_i \mathbf{a_i}=\sum_{i=1}^3 z_i \mathbf{d_i}\in \mathbb{E}^3$ we indicate the coordinate tuples for the rotating outer basis by $\mathsf{\breve z}=(\breve z_1, \breve z_2, \breve z_3)\in \Real^3$ and for the director basis by $\mathsf{z}=(z_1,z_2,z_3)\in \Real^3$. The director basis can be transformed into the rotating outer basis by the tensor-valued rotation $\mathbf{R}$, i.e.\ $\mathbf{R}=\mathbf{a_i}\otimes \mathbf{d_i} =R_{ij} \mathbf{a_i}\otimes \mathbf{a_j}\in \mathbb{E}^3\otimes \mathbb{E}^3$ with associated orthogonal matrix $\mathsf{R}=(R_{ij})=(\mathbf{d_i}\cdot\mathbf{ a_j})\in SO(3)$. For the coordinate tuples, $\mathsf{z}=\mathsf{R} \cdot \mathsf{\breve z}$ holds. The cross-product $\mathsf{a}\times \mathsf{A}\in \Real^{3\times3}$ between a vector $\mathsf{a}\in \Real^3$ and a matrix $\mathsf{A}\in \Real^{3\times3}$ is defined by $(\mathsf{a}\times \mathsf{ A})\cdot \mathsf{z}=\mathsf{ a}\times(\mathsf{A\cdot z})$ for all $\mathsf{z}\in \Real^3$. Moreover, we abbreviate $\mathsf{P}_k=\mathrm{diag}(1,1,k)$, $k\in \mathbb{R}$ and $\mathsf{e_i}\in \mathbb{R}^3$, $i=1,2,3$  for the canonical basis tuples.
\end{notation}

\subsubsection*{Set-up A: inflow in Lagrangian parameterization} Let $\mathcal{Q}_T=\{ (\sigma,t) \in \mathbb{R}^2 \,|\, \sigma \in (-\ell(t),0), \, \ell(t)=t, \, t \in (0,T] \}$ be the flow domain enlarging over time where $\mathcal{Q}_0=\emptyset$ holds initially for $t=0$. The rod model for the inflow set-up (i.e.\ inflow at the nozzle $\sigma=-\ell(t)$ and (stress-)free jet end $\sigma=0$) reads
\begin{align}\label{eq:L_inflow} 
\mathsf{R} \cdot \partial_t \mathsf{\breve{r}} &= \mathsf{v} \\ \nonumber
\partial_t \mathsf{R} &= -\mathsf{\omega} \times \mathsf{R}\\ \nonumber
\partial_t e {\sf e_3}&= 
\partial_\sigma \mathsf{v} + \mathsf{\kappa}\times \mathsf{v} + e\mathsf{e_3}\times \mathsf{\omega}\\ \nonumber
\partial_t \mathsf{\kappa} & = 
\partial_\sigma \mathsf{\omega} + \mathsf{\kappa}\times \mathsf{\omega}\\ \nonumber
\partial_t \mathsf{v} &=
\frac{1}{\mathrm{Re}}\left(\partial_\sigma \mathsf{n} + \mathsf{\kappa}\times \mathsf{n}\right) 
+ \mathsf{v} \times \mathsf{\omega} + \frac{1}{\mathrm{Fr}^2}\mathsf{R}\cdot\mathsf{e_g}+\mathsf{k_\Omega}\\ \nonumber
\mathsf{P_2} \cdot \partial_t \frac{\mathsf{\omega}}{e}  &= 
\frac{4}{\mathrm{Re}}(\partial_\sigma \mathsf{m} + \mathsf{\kappa}\times \mathsf{m})
+ \frac{16}{\epsilon^2 \mathrm{Re}} e\mathsf{e_3}\times\mathsf{n}+\mathsf{l_\Omega}
\end{align}
with Coriolis and centrifugal forces as well as corresponding couples due to the rotating reference frame
\begin{align*}
\mathsf{k_\Omega}&= -\frac{2}{\mathrm{Rb}} \mathsf{R}\cdot \mathsf{e_\Omega}\times \mathsf{v}
-\frac{1}{\mathrm{Rb}^2} \mathsf{R}\cdot\left(\mathsf{e_\Omega}\times (\mathsf{e_\Omega}\times \mathsf{\breve{r}})\right)\\
\mathsf{l_\Omega}&=  \mathsf{P_2}\cdot \frac{1}{e}\left(\mathsf{\omega}+\frac{1}{\mathrm{Rb}}\mathsf{R}\cdot \mathsf{e_\Omega}\right) \times \left(\mathsf{\omega}+\frac{1}{\mathrm{Rb}}\mathsf{R}\cdot \mathsf{e_\Omega}\right)
+ \mathsf{P_2}\cdot \left[ \left(\frac{\mathsf{\omega}}{e}\times \frac{1}{\mathrm{Rb}}\mathsf{R}\cdot \mathsf{e_\Omega}\right)
+ \frac{1}{\mathrm{Rb}} \frac{\partial_t e}{e^2} \mathsf{R}\cdot \mathsf{e_\Omega}  \right]
\end{align*}
and material laws
\begin{align*}
n_3=3 \frac{1}{e^2}(\partial_\sigma v_3+\kappa_1 v_2-\kappa_2 v_1), \qquad   \mathsf{m}=\frac{3}{4}\frac{1}{e^3}\mathsf{P}_{2/3}\cdot (\partial_\sigma \omega +\kappa\times \omega)\,.
\end{align*}
The material laws can be alternatively expressed in terms of the strain rates $\partial_t e$ and $\partial_t \kappa$, since $\partial_t e=\partial_\sigma v_3+\kappa_1 v_2-\kappa_2 v_1$ and $\partial_t \kappa=\partial_\sigma \omega +\kappa\times \omega$. In particular, they are linear in the strain rates. However, to avoid mixed time-space derivatives when plugging the material laws into the balance equations we use the stated spatial representation yielding second spatial derivatives in the dynamic equations. A strict classification of the whole system \eqref{eq:L_inflow} is not possible, but it has a hyperbolic-parabolic character with ordinary differential equations for curve and rotation group (director triad). The boundary conditions are
\begin{align*}
\mathsf{\breve{r}}(-\ell(t),t)&=\mathsf{e_2}, \qquad  
\mathsf{R}(-\ell(t),t)=\mathsf{e_1}\otimes \mathsf{e_1} - \mathsf{e_2}\otimes \mathsf{e_3} + \mathsf{e_3}\otimes \mathsf{e_2} \\ 
e(-\ell(t),t)&=1, \qquad \,\,
\mathsf{\kappa}(-\ell(t),t)=\mathsf{0}, \qquad 
\mathsf{v}(-\ell(t),t)=\mathsf{e_3}, \qquad 
\mathsf{\omega}(-\ell(t),t)=\mathsf{0}\\
\mathsf{n}(0,t)&=\mathsf{0}, \qquad \,\,
\mathsf{m}(0,t)=\mathsf{0}.
\end{align*}

\subsubsection*{Set-up B: inflow-outflow in Eulerian parameterization} Let $\mathcal{S}_T=\{(s,t)\in \mathbb{R}^2\,|\,s\in (0,\ell), \,t\in (0,T],\, \ell >0 \text{ fixed}\}$ be the flow domain fixed over time. The rod model for the inflow-outflow set-up (i.e.\ inflow at the nozzle $s=0$ and outflow at a prescribed length $s=\ell$) reads
\begin{align}\label{eq:E} 
\mathsf{R} \cdot \partial_t \mathsf{\breve{r}} &=
\mathsf{v} -u \mathsf{e_3}\\ \nonumber
\partial_t \mathsf{R} &= -(\mathsf{\omega}- u\mathsf{\kappa}) \times \mathsf{R}\\ \nonumber
\partial_s (u\mathsf{e_3}) & = 
\partial_s \mathsf{v} + \mathsf{\kappa}\times \mathsf{v} +
\mathsf{e_3}\times \mathsf{\omega}\\ \nonumber
\partial_t \mathsf{\kappa} + \partial_s (u \mathsf{\kappa}) & = 
\partial_s \mathsf{\omega} + \mathsf{\kappa}\times \mathsf{\omega}\\  \nonumber
\partial_t A + \partial_s (u A) & =  0\\ \nonumber
\partial_t(A \mathsf{v}) + \partial_s(u A \mathsf{v}) &=
\frac{1}{\mathrm{Re}} \left(\partial_s \mathsf{n} + \mathsf{\kappa}\times \mathsf{n}\right) 
+ A \mathsf{v} \times \mathsf{\omega} + \frac{1}{\mathrm{Fr}^2}A\mathsf{R}\cdot \mathsf{e_g}+\mathsf{k_\Omega}\\ \nonumber
\mathsf{P_2}\cdot (\partial_t (A^2\mathsf{\omega}) + \partial_s(u A^2\mathsf{\omega})) &=
\frac{4}{\mathrm{Re}}(\partial_s \mathsf{m} + \mathsf{\kappa}\times \mathsf{m})
+ \frac{16}{\epsilon^2 \mathrm{Re}} \mathsf{e_3}\times \mathsf{n}+\mathsf{l_\Omega}
\end{align}
with
\begin{align*}
\mathsf{k_\Omega}&=-\frac{2}{\mathrm{Rb}}\mathsf{R}\cdot \mathsf{e_\Omega}\times A\mathsf{v} - \frac{1}{\mathrm{Rb}^2}A \mathsf{R}\cdot(\mathsf{e_\Omega} \times(\mathsf{e_\Omega} \times \mathsf{\breve r})) \\ 
\mathsf{l_\Omega}&= \mathsf{P_2} \cdot A^2\left(\mathsf{\omega}+\frac{1}{\mathrm{Rb}}\mathsf{R}\cdot \mathsf{e_\Omega}\right)\times \left(\mathsf{\omega}+\frac{1}{\mathrm{Rb}}\mathsf{R}\cdot \mathsf{e_\Omega}\right) +  
\mathsf{P_2}\cdot\left[\left(A^2\mathsf{\omega}\times \frac{1}{\mathrm{Rb}}\mathsf{R}\cdot \mathsf{e_\Omega}\right) + \frac{1}{\mathrm{Rb}} A^2 \partial_s u \mathsf{R}\cdot \mathsf{e_\Omega} \right] 
\end{align*}
and material laws
\begin{align*}
n_3=3 A \partial_s u, \qquad  \mathsf{m}=\frac{3}{4} A^2 \mathsf{P}_{2/3}\cdot (\partial_s \mathsf{\omega} + \mathsf{\kappa}\times \mathsf{\omega})
\end{align*}
The boundary conditions for $t\in[0,T]$ are
\begin{align*}
\mathsf{\breve{r}}(0,t)&=\mathsf{e_2}, \qquad  
\mathsf{R}(0,t)=\mathsf{e_1}\otimes \mathsf{e_1} - \mathsf{e_2}\otimes \mathsf{e_3} + \mathsf{e_3}\otimes \mathsf{e_2} \\ 
u(0,t)&=1, \qquad \,\,
\mathsf{\kappa}(0,t)=\mathsf{0}, \qquad 
\mathsf{v}(0,t)=\mathsf{e_3}, \qquad 
\mathsf{\omega}(0,t)=\mathsf{0}, \qquad
A(0,t)=1\\
\mathsf{n}(\ell,t)&=\mathsf{0}, \qquad \,\,
\mathsf{m}(\ell,t)=\mathsf{0}.
\end{align*}
Appropriate initial conditions are specified later on. \\

The computation of the stated model equations \eqref{eq:L_inflow}, \eqref{eq:E} from the general invariant system \eqref{eq:L_invariant} is straightforward, but lengthy. For more details about the determination we refer to \cite{arne:p:2010}. The systems can be easily simplified to 2d. Note that the dimension plays no role for the development of the numerical scheme but the reduction to 2d will be used for the simulation of a bench-mark test scenario in Section~\ref{sec:4} (cf.\ Fig.~\ref{fig:1} and Eqs.~\eqref{eq:L_2d}, \eqref{eq:E_2d} for 2d rotational spinning under neglect of gravity). 

\begin{remark}[Unit quaternions for rotations]
The rotations $\mathsf{R}\in SO(3)$ can be parameterized, e.g.\ in Euler angles or unit quaternions \cite{mahadevan:p:1996}. We use unit quaternions since this variant offers a very elegant way of formulating and computing the evolution equation for $\mathsf{R}$ (second equation of \eqref{eq:L_inflow} or \eqref{eq:E} respectively). Define 
\begin{align*}
\mathsf{R}(\mathsf{q})=
 \left( \begin{array}{ccc}
q_1^2-q_2^2-q_3^2+q_0^2 & 2 (q_1 q_2 - q_0 q_3) & 2 (q_1 q_3 + q_0 q_2)\\
2 (q_1 q_2 + q_0 q_3)  & -q_1^2+q_2^2-q_3^2+q_0^2 & 2 (q_2 q_3 - q_0 q_1)\\
2 (q_1 q_3 - q_0 q_2) & 2 (q_2 q_3 + q_0 q_1) & -q_1^2-q_2^2+q_3^2+q_0^2 
\end{array} \right), 
\end{align*}
with unit quaternions $\mathsf{q}=(q_0,q_1,q_2,q_3)$, $\|\mathsf{q}\|=1$, then $\partial_t \mathsf{R} = -\mathsf{\omega} \times \mathsf{R}$ becomes
$ \partial_t\mathsf{q}=\mathcal{A}(\mathsf{\omega})\cdot \mathsf{q}$ (or respectively, $\partial_t \mathsf{R} = -(\mathsf{\omega}- u\mathsf{\kappa}) \times \mathsf{R}$ becomes $ \partial_t\mathsf{q}=\mathcal{A}(\mathsf{\omega}-u\kappa)\cdot \mathsf{q}$)
with skew-symmetric matrix 
\begin{align*}
\mathcal{A}(\mathsf{z}) = \frac{1}{2}
\left( \begin{array}{cccc}
0 & z_1 & z_2 & z_3\\
-z_1 & 0 & z_3 & -z_2\\
-z_2 & -z_3 & 0 & z_1\\
-z_3 & z_2 & -z_1 & 0
\end{array} \right).
\end{align*}
\end{remark}

%%%%%%%%%%%%%%%%%%%%%%%%%%%%%%%%%%%%%%%%%%%%%%%%%%%%%%%%%%

\section{Numerical scheme}\label{sec:3}

Finite volume schemes are well-established for the numerical solution of time-dependent partial differential equations for various applications \cite{versteeg:b:2007}. In the following we focus on the inflow problem in the Lagrangian parameterization because of the tricky initialization and the handling of the length change $\ell(t)$. The inflow-outflow problem in the Eulerian parameterization where the domain length is fixed is comparatively much easier. A respective scheme can be established straightforward in an accordant way (cf.\ \cite{arne:p:2012} for the 2d scenario). To set up the numerical concept for our inflow problem we rewrite \eqref{eq:L_inflow} in a more convenient formulation, for this purpose we define $\mathsf{0}_{k}$ as the zero vector in $\mathbb{R}^k$. We introduce the vector of unknowns 
\begin{displaymath}
\phi=\left(n_1, n_2, e, \mathsf{\breve{r}}, \mathsf{q}, \kappa, \mathsf{v}, \varpi \right) \in\R^{19}
\end{displaymath}
with $\varpi={\omega}/{e}$.
To take account of the differential-algebraic structure of the underlying model, we additionally consider the mapping 
\begin{displaymath}
\mathsf{z}(\phi)= \left(\mathsf{0}_2, e, \mathsf{\breve r}, \mathsf{q}, \kappa, \mathsf{v}, \varpi\right) \in\R^{19}
\end{displaymath}
that consists of all variables possessing an evolution equation in \eqref{eq:L_inflow}. Finite volume schemes are based on the integral form of the governing equations that are expressed in terms of flux functions and source terms. Therefore, we summarize the constituents with respect to their physical meaning and later used numerical approximation. The upper index $u,d,c$ indicates the respective fluxes considered for up-, down-winded and central differences:
\begin{align*}
\mathsf{f}^u(\phi)& =
\left(\mathsf{v}, \mathsf{0}_{7}, e\varpi, \mathsf{0}_2, \frac{3}{\mathrm{Re}}\frac{1}{e^2} (\kappa_1v_2-\kappa_2v_1), 
\frac{3}{\mathrm{Re}}\frac{1}{e^2}\mathsf{P}_{1/3}\cdot (\kappa\times\varpi) \right)\\ 
\mathsf{f}^d(\phi)& =
\left(\mathsf{0}_{13}, \frac{1}{\mathrm{Re}}n_1, \frac{1}{\mathrm{Re}}n_2, \mathsf{0}_4\right)\\
\mathsf{f}^c(\phi,\partial_\sigma \mathsf{h}(\phi)) & =
\left(\mathsf{0}_{15}, \frac{3}{\mathrm{Re}}\frac{1}{e^2} \partial_{\sigma} v_3, 
\frac{3}{\mathrm{Re}}\frac{1}{e^3}\mathsf{P}_{1/3}\cdot \partial_{\sigma}\left(e\varpi \right) \right)
\intertext{and}
\mathsf{p}(\phi,\partial_\sigma \mathsf{g}(\phi)) & =
\left(\mathsf{0}_{13}, \frac{3}{\mathrm{Re}}\frac{\kappa}{e}\times \left(0, 0, \partial_\sigma v_3\right), 
\frac{3}{\mathrm{Re}}\frac{1}{e^3}\mathsf{P}_{1/2}\cdot\left(\kappa\times(\mathsf{P}_{2/3}\cdot\partial_\sigma\left(e\varpi\right))\right)   \right) \, . 
\end{align*}
Here, $\mathsf{h}(\phi)=\left(\mathsf{0}_{15}, v_3, e\varpi \right)$ and 
$\mathsf{g}(\phi)=\left(\mathsf{0}_{13}, v_3, v_3, 0, e\varpi \right)$ hold.
The remaining source terms are collected in $\mathsf{q}(\phi)$. Due to this dispartment, the system \eqref{eq:L_inflow} becomes
\begin{align}\label{gov_eq}
\partial_t \mathsf{z}(\phi)
=\partial_\sigma\mathsf{f}^u(\phi) + \partial_\sigma\mathsf{f}^d(\phi)
+\partial_\sigma \mathsf{f}^c(\phi,\partial_\sigma \mathsf{h}(\phi))+ \mathsf{p}(\phi,\partial_\sigma \mathsf{g}(\phi))+\mathsf{q}(\phi)
\end{align}
where the closure relations are incorporated.

\begin{figure}[tb]
\begin{center} \hspace*{-1cm}
\psfrag{lt}{$\ell(t)$}
\psfrag{lt1}{$\ell(t+\triangle t)$}
\psfrag{1}{$\sigma_1$}
\psfrag{2}{$\sigma_2$}
\psfrag{N}{$\sigma_N$} 
\psfrag{Nt}{$N=N(t)$}
\psfrag{Nt1}{$N=N(t+\triangle t)$}
\psfrag{0}{$\sigma_0$}
\psfrag{g}{initialization}
\psfrag{Noz}{nozzle}
\psfrag{End}{jet end}
\psfrag{05}{$\sigma_{1/2}$}
\psfrag{15}{$\sigma_{3/2}$}
\psfrag{N2}{$\sigma_{N+1/2}=0$}
\includegraphics[scale=0.7]{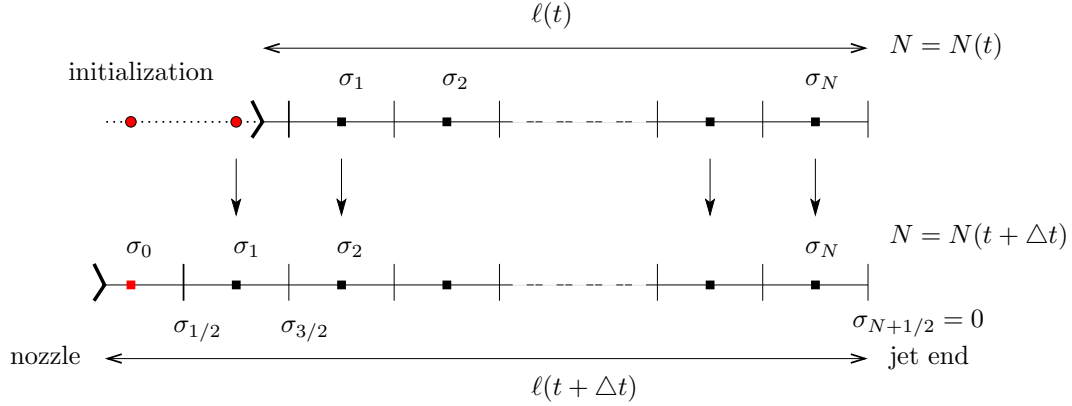}
\caption{\label{fig:skizze} Spatial discretization of the growing jet with $N(t)$ equally sized dynamic cells (centers are marked by black squares). The cell edge $\sigma_{N+1/2}=\sigma=0$ represents the jet end; the complete cell $[\sigma_{1/2},\sigma_{3/2}]$ is the closest dynamic one to the nozzle at $\sigma=-\ell(t)$. For the initialization static cells are introduced before/around the nozzle (red circles), whose quantities are given by the nozzle conditions.}
\end{center}
\end{figure}

Concerning the space discretization we introduce a constant cell size $\triangle \sigma$ and define the number $N(t)$ of dynamic cells for the time-dependent jet length $\ell(t)$ by help of the floor function $\lfloor.\rfloor$, 
\begin{align*}
N(t)=\left \lfloor \frac{\ell(t)}{\triangle \sigma} \right \rfloor, \quad \quad \quad \quad \quad \sigma_{(j+1)/2}=-\left(N(t)-\frac{j}{2}\right)\triangle \sigma, \quad j=0,\ldots,2N(t),
\end{align*}
where $\sigma_i$, $i=1,\ldots, N(t)$ denote the cell centers, cf.\ Fig.~\ref{fig:skizze}. The cell edge $\sigma_{N+1/2}=\sigma=0$ represents the jet end, and $[\sigma_{1/2},\sigma_{3/2}]$ is the closest dynamic cell to the nozzle located at $\sigma=-\ell(t)=-t$.
The jet growth is realized by adding new cells at the nozzle, hereby the dynamics of the physical quantities is not considered till the cells have completely left the nozzle and become dynamic. Before they are treated as static and initialized by the boundary condition at the nozzle. The number of static cells depends on the ongoing length increase in the time interval under consideration $[t,t+\triangle t]$, we use $M(t)=N(t+\triangle t)-N(t)$. The introduction of the static cells before/around the nozzle allows the adequate initialization of a jet of length $\ell(t)<\triangle \sigma$ and a stable numerical treatment of the temporal evolution.

The idea is now to integrate \eqref{gov_eq} over the control volume / cell $[\sigma_{i-1/2},\sigma_{i+1/2}]$, $i=1,\ldots,N(t)$ and to set up a differential algebraic system (DAE) in time for the cell averages $\phi_i$ of the unknown quantities 
\begin{align*}
\phi_i(t):=\frac{1}{\triangle \sigma}\int_{\sigma_{i-1/2}}^{\sigma_{i+1/2}}\phi(\sigma,t)\, \mathrm{d}\sigma, \quad i=1,\ldots,N(t).
\end{align*}
The resulting DAE with $\mathsf{z}_i(t)=\mathsf{z}(\phi_i(t))$ has the form
\begin{align}\label{weak}
\frac{\mathrm{d}}{\mathrm{d}t}
\mathsf{z}_i &= \frac{1}{\triangle \sigma}
[(\mathsf{f}^u_{i+1/2}-\mathsf{f}^u_{i-1/2})
+(\mathsf{f}^d_{i+1/2}-\mathsf{f}^d_{i-1/2})
+(\mathsf{f}^c_{i+1/2}-\mathsf{f}^c_{i-1/2})]\nonumber\\ 
&\quad+\frac{1}{\triangle \sigma}\int_{\sigma_{i-1/2}}^{\sigma_{i+1/2}}\mathsf{p}(\phi,\partial_{\sigma}\mathsf{g}(\phi))\, \mathrm{d}\sigma
+\frac{1}{\triangle \sigma}\int_{\sigma_{i-1/2}}^{\sigma_{i+1/2}}\mathsf{q}(\phi)\, \mathrm{d}\sigma.
\end{align}
To express all constituents in terms of the time-dependent $\phi_i(t)$, we define numerical flux functions $H^u$, $H^d$, $H^c$ in an up-, down-winded and central manner according to the behavior of the physical fluxes, i.e.\ we apply the upwind-strategy for the convective terms, the downwind-strategy for the normal forces and the central approximation for the viscous parts,
\begin{align*}
\mathsf{f}^u_{i+1/2}\approx H^u(\phi_i,\phi_{i+1})&:=\mathsf{f}^u(\phi_i)\\
\mathsf{f}^d_{i+1/2}\approx H^d(\phi_i,\phi_{i+1})&:=\mathsf{f}^d(\phi_{i+1})\\ 
\mathsf{f}^c_{i+1/2}\approx H^c(\phi_i,\phi_{i+1})&:=\mathsf{f}^c\left(\frac{\phi_i+\phi_{i+1}}{2},\frac{\mathsf{h}(\phi_{i+1})-\mathsf{h}(\phi_i)}{\triangle\sigma}\right),
\end{align*}
$i=1,\ldots,N(t)-1$. The integrals that contain the source terms are approximated by means of
\begin{align*}
\frac{1}{\triangle \sigma}\int_{\sigma_{i-1/2}}^{\sigma_{i+1/2}}\mathsf{p}(\phi,\partial_\sigma{\mathsf{g}}(\phi))\, \mathrm{d}\sigma 
&\approx P(\phi_{i-1},\phi_i):=\mathsf{p}\left(\phi_i,\frac{\mathsf{g}(\phi_i)-\mathsf{g}(\phi_{i-1})}{\triangle \sigma} \right), && i=2,\ldots,N(t)\\
\frac{1}{\triangle \sigma}\int_{\sigma_{i-1/2}}^{\sigma_{i+1/2}}\mathsf{q}(\phi)\, \mathrm{d}\sigma
&\approx \mathsf{q}(\phi_i), && i=1,\ldots,N(t).
\end{align*}
As for the boundaries at nozzle and stress-free jet end, the proposed discretizations make use of the respective boundary conditions -- collected in $\phi_{noz}$ and $\phi_{end}$ -- in a natural way. We use
\begin{align*}
\mathsf{f}_{1/2}^u\approx \mathsf{f}^u(\phi_{noz}), \hspace*{0.6cm} \mathsf{f}_{1/2}^d&\approx \mathsf{f}^d(\phi_1),\hspace*{1.1cm} 
\mathsf{f}_{1/2}^c= \mathsf{f}^c\left(\phi_{noz},\frac{\mathsf{h}(\phi_1)-\mathsf{h}(\phi_{noz})}{\triangle \sigma}\right) &&\text{ at nozzle}\\
\mathsf{f}_{N+1/2}^u\approx \mathsf{f}^u(\phi_N), \quad \mathsf{f}_{N+1/2}^d&\approx \mathsf{f}^d(\phi_{end}), \quad 
\mathsf{f}_{N+1/2}^c=\mathsf{0} && \text{ at jet end}
\end{align*}
In accordance we take 
$\int_{\sigma_{1/2}}^{\sigma_{3/2}}\mathsf{p}(\phi,\partial_{\sigma}\mathsf{g}(\phi))\, \mathrm{d}\sigma/\triangle \sigma \approx P(\phi_{noz},\phi_1)$
for the source term in the first control volume. 
Inserting the numerical flux functions $H=H^u+H^d+H^c$ and source term discretizations into system \eqref{weak}, we finally obtain its semi-discrete analogon 
\begin{align}\label{stern}
\frac{\mathrm{d}}{\mathrm{d}t}\mathsf{z}_i=\frac{1}{\triangle \sigma}
(H(\phi_i,\phi_{i+1})-H(\phi_{i-1},\phi_{i}))+P(\phi_{i-1},\phi_i)+\mathsf{q}(\phi_i).
\end{align}

The system of DAEs \eqref{stern} is of index 2 (according to the definition of \cite{hairer:b:1996}). For the time integration stiff accurate implicit Runge-Kutta schemes, e.g.\ Radau IIa methods, are suitable, cf.\ Remark~\ref{rem:1}. We employ a constant time step $\triangle t$. The resulting nonlinear system of equations is solved with a Newton method. 

\begin{remark}[Runge-Kutta methods for DAEs]\label{rem:1}
Consider an autonomous differential-algebraic system $\mathrm{d}u/\mathrm{d}t=f(u,v)$, $0=g(u,v)$ with initial value $(u,v)(t_0)=(u_0,v_0)$. Let a time discretization $t_0 <t_1<\ldots < t_M$ with step size $\triangle t_n=t_{n+1}-t_n$ be given. An appropriate implicit Runge-Kutta scheme of level $s$ (with coefficients $\mathbf{A}=(a_{ij})\in\mathbb{R}^{s\times s}$, $\mathbf{b}=(b_i)\in\mathbb{R}^s$) has the form
\begin{align*}
u_{n+1}=u_n+\triangle t_n\sum_{j=1}^s b_j f(k_j,l_j), \quad \quad \quad v_{n+1}=v_{n}+(l_1-v_n|\ldots|l_s-v_n)\mathbf{A}^{-T}\mathbf{b}, 
\end{align*}
where the levels $k_i$, $l_i$ are the solutions of the nonlinear system 
\begin{align*}
k_i=u_n+\triangle t_n\sum_{j=1}^n a_{ij} f(k_j,l_j), \quad \quad \quad 0=g(k_i,l_i), \quad \quad \quad i=1,\ldots, s.
\end{align*}
A scheme is called stiff accurate if the coefficients satisfy $b_i=a_{si}$. Radau IIa methods possess this property, for $s=1,2$ they are specified by the following Butcher arrays for the coefficients: 
\begin{align*}
\begin{array}{c| c} \mathbf{c} & \mathbf{A}\\ \hline\\[-2.4ex] &\mathbf{b}^T \end{array} \quad \quad \quad \quad \quad
\begin{array}{c| c} 1 & 1 \\ \hline & 1 \end{array} \quad \quad \quad
\begin{array}{c| c c} 1/3 & 5/12 & -1/12 \\ 1 & 3/4 & 1/4 \\ \hline & 3/4 & 1/4 \end{array}\,.
\end{align*}
Note that $c_i=\sum_{j} a_{ij}$ as the Runge-Kutta scheme is invariant with respect to autonomization.
The Radau IIa method for $s=1$ is well-known as implicit Euler method. For details on Runge-Kutta methods for DAEs see e.g.\ \cite{hairer:b:1989}.
\end{remark}

\begin{figure}[b]
\includegraphics[scale=0.425]{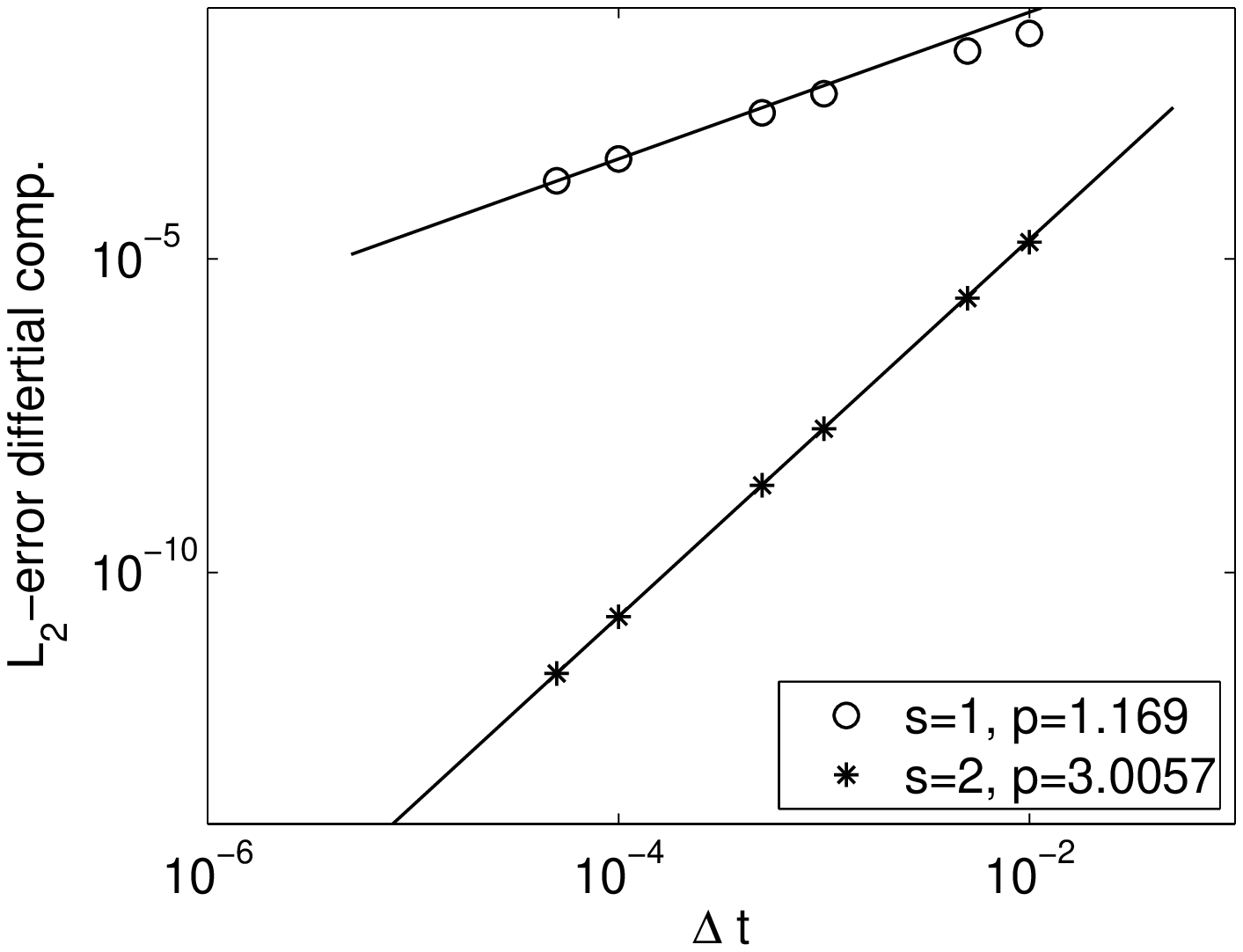}
\includegraphics[scale=0.425]{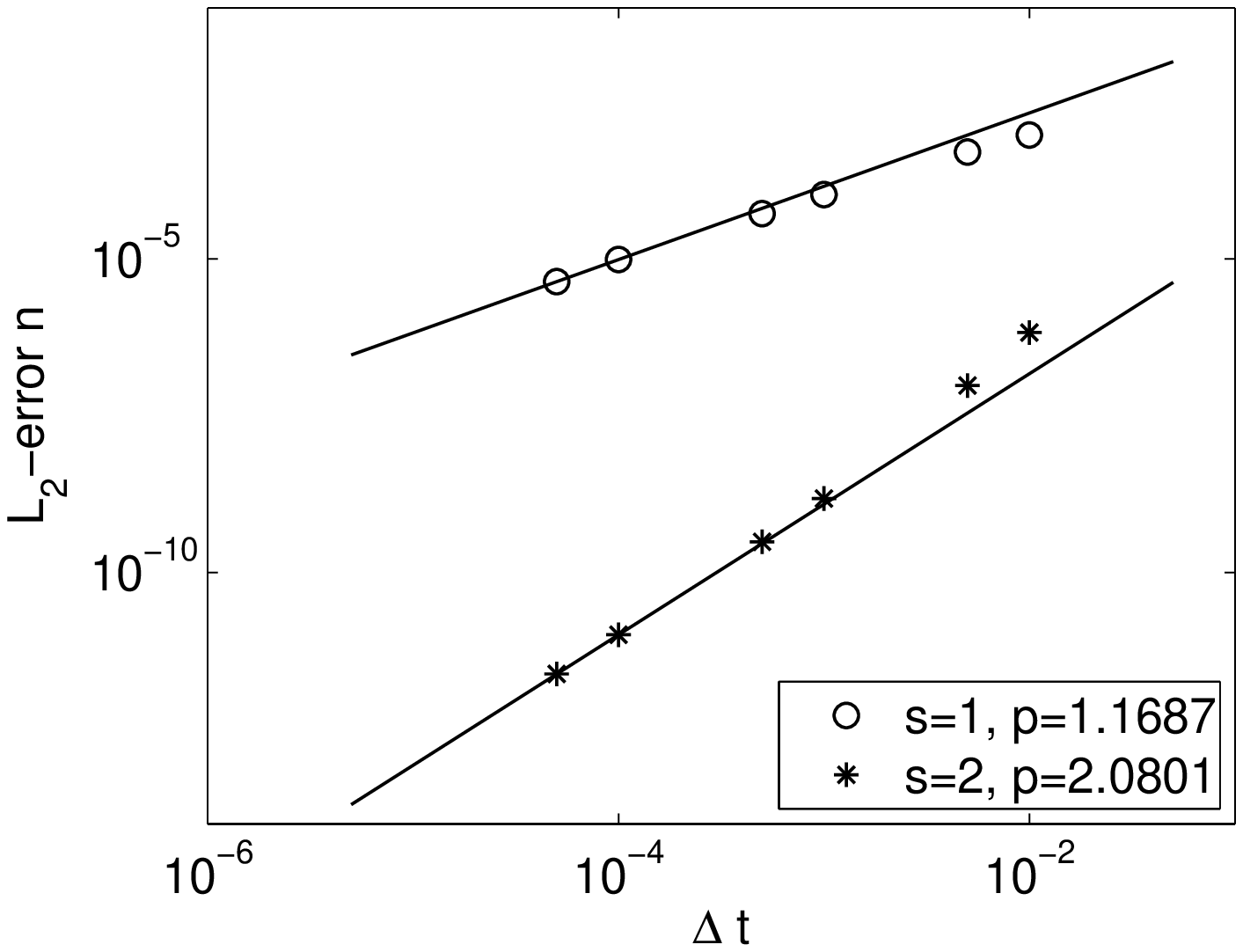}\\
\includegraphics[scale=0.425]{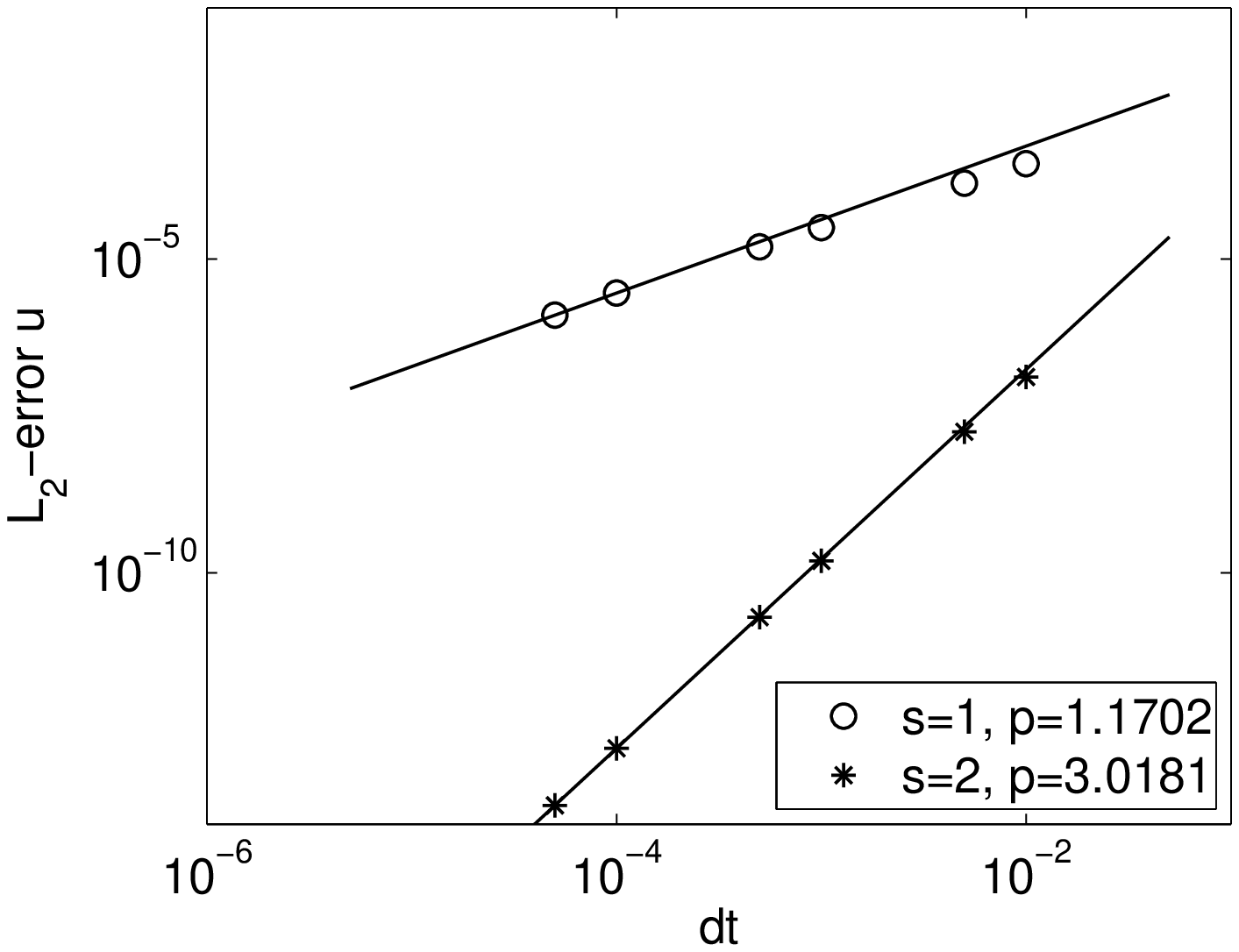}\
\caption{\label{fig:conv_euler} Convergence of Radau IIa methods for DAEs of semi-discretized inflow-outflow problem in Eulerian parameterization  -- in consistence with theory. Absolute $L^2(0,\ell)$-error for fixed end time $T$. {\it Left:} differential variables. {\it Right:} algebraic variables $(\mathsf{n})$. {\it Bottom:} The intrinsic velocity $u$ that has a special role in \eqref{eq:E} on first glance behaves like all the other differential variables.}
\end{figure}

\begin{figure}[b]
\includegraphics[scale=0.425]{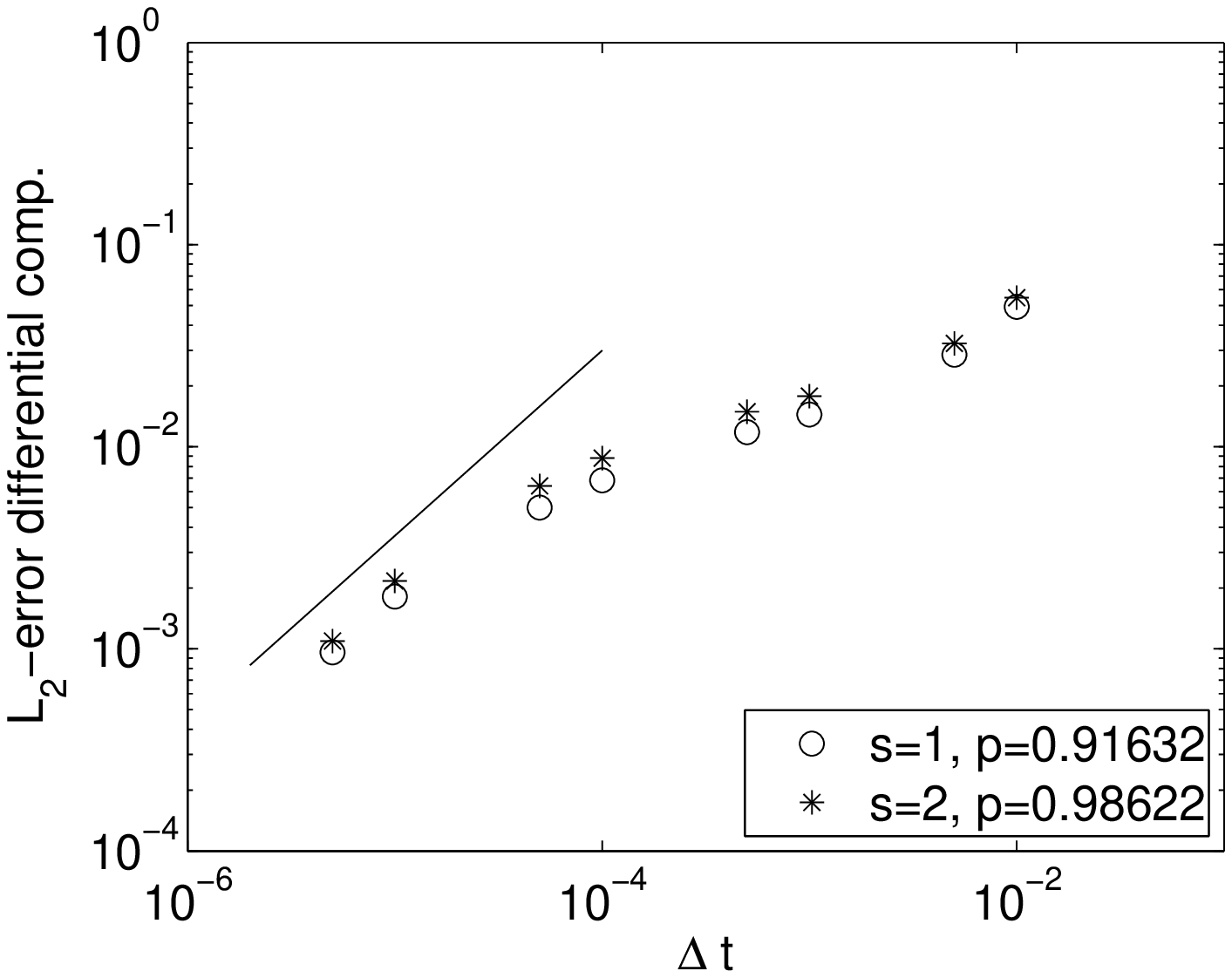}
\includegraphics[scale=0.425]{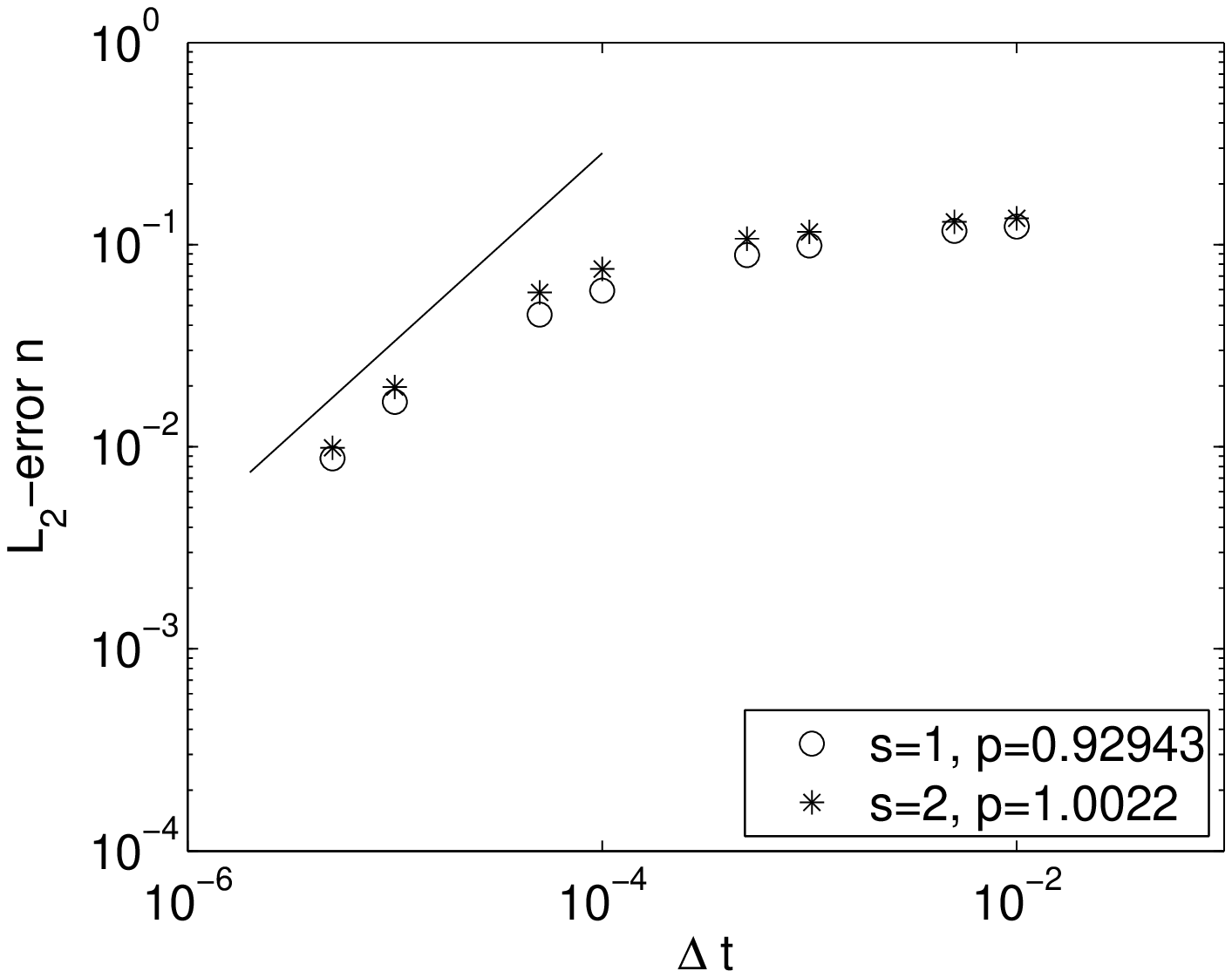}
\caption{\label{fig:conv_lagrange} Convergence of Radau IIa methods for DAEs \eqref{stern} corresponding to inflow problem in Lagrangian parameterization with enlarging domain and free end. Absolute $L^2(0,\ell(T))$-error for fixed end time $T$. {\it Left:} differential variables. {\it Right:} algebraic variables. Straight line with slope 1 indicates first order convergence.}
\end{figure}

The space discretization via the finite volume scheme is of first order convergence. We aim at an appropriate, stiff accurate time discretization. In the context of differential-algebraic equations, the order of convergence $p$ depends on the index and the Runge-Kutta methods often tend to loose an order for the variables associated to the algebraic equations. Considering a DAE with index 2, the Radau IIa method with $s=2$ has $p=3$ for the differential variables and $p=2$ for the algebraic ones, whereas the implicit Euler method has $p=1$ for both kind of variables, \cite{hairer:b:1989}. This theoretical result is confirmed by our numerical tests for the inflow-outflow problem in the Eulerian parameterization on the time-independent space interval $[0,\ell]$ (using an fixed equidistant spatial grid with $\triangle s=\ell/N$ where nozzle and outflow are located at the cell edges $s=s_{1/2}=0$ and $s=s_{N+1/2}=\ell$, respectively), see Figure~\ref{fig:conv_euler}. For the inflow problem in the Lagrangian parameterization, the jet length $\ell(t)$ and hence the space discretization is time-dependent. There is no strict separation of space and time as in an usual semi-discretization. Therefore, it is not surprising that the performance of the Radau IIa method with $s=2$ differs to the theoretical result. We observe a loss of convergence order due to the chosen discretization / initialization at the nozzle boundary. In correspondence to the space discretization we obtain here first order convergence in time for both Radau variants, Figure~\ref{fig:conv_lagrange}.
Consequently, to obtain higher convergence for the inflow problem the spatial scheme needs to be modified. 

\begin{remark}[Choice of temporal and spatial grid sizes]\label{rem:grid}
For the forthcoming numerical simulations of the inflow-outflow problem in the Eulerian parameterization the choice of time step and spatial grid size follows the CFL-condition with respect to the intrinsic velocity. In the inflow problem in the Lagrangian parameterization $\triangle t$ and $\triangle \sigma$ do not need coercively to be coupled, but it turns out that they have to be adapted in view of the parameters ($\mathrm{Re}$,$\mathrm{Rb}$,$\mathrm{Fr}$) of the problem. Smaller parameters imply in general faster and larger changes in the jet dynamics which require a finer resolution. Otherwise it might happen that the used Newton method does not converge.
\end{remark}

%%%%%%%%%%%%%%%%%%%%%%%%%%%%%%%%%%%%%%%%%%

\section{Simulation results}\label{sec:4}

\begin{figure}[H]
\includegraphics[scale=0.425]{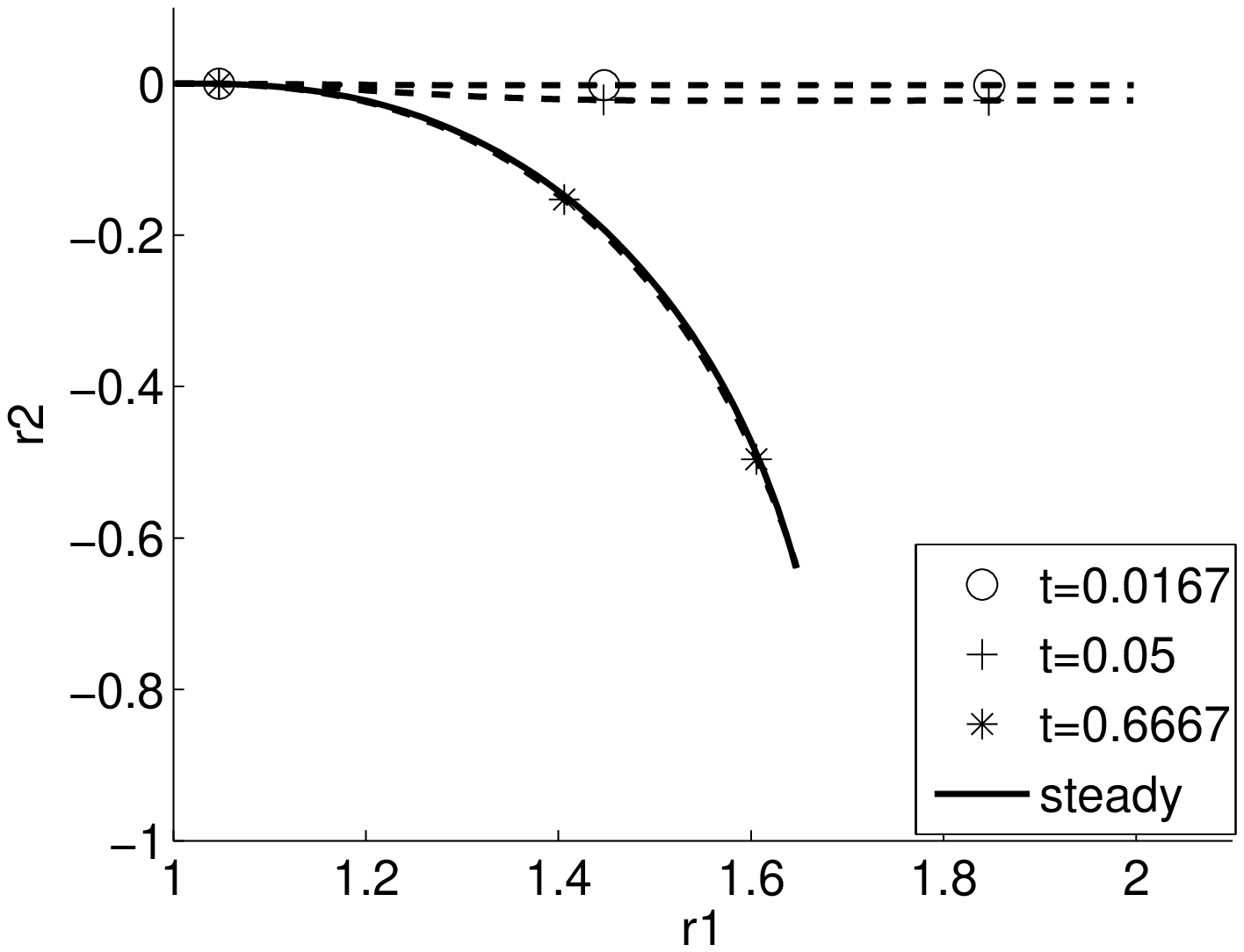}
\includegraphics[scale=0.425]{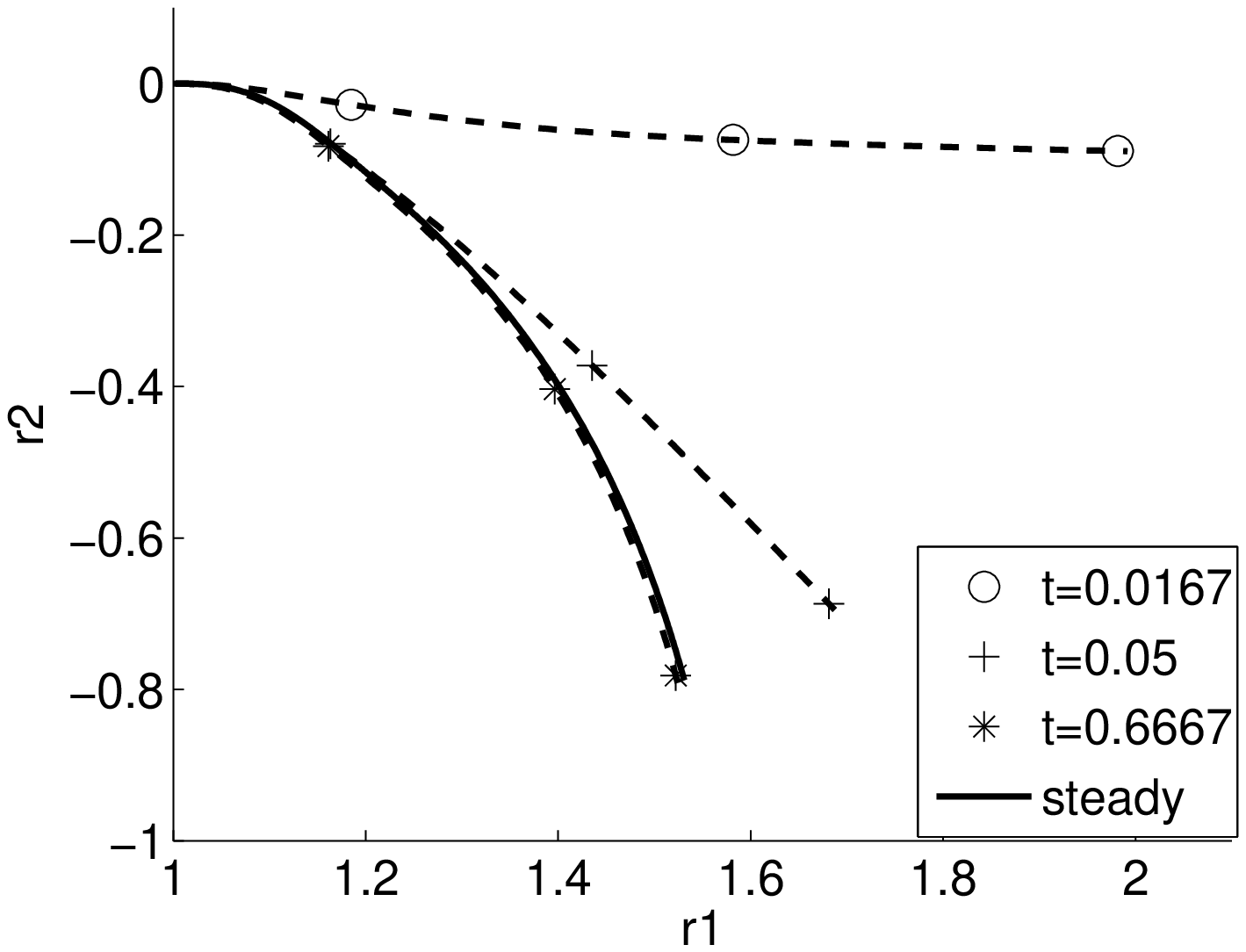}\\
\includegraphics[scale=0.425]{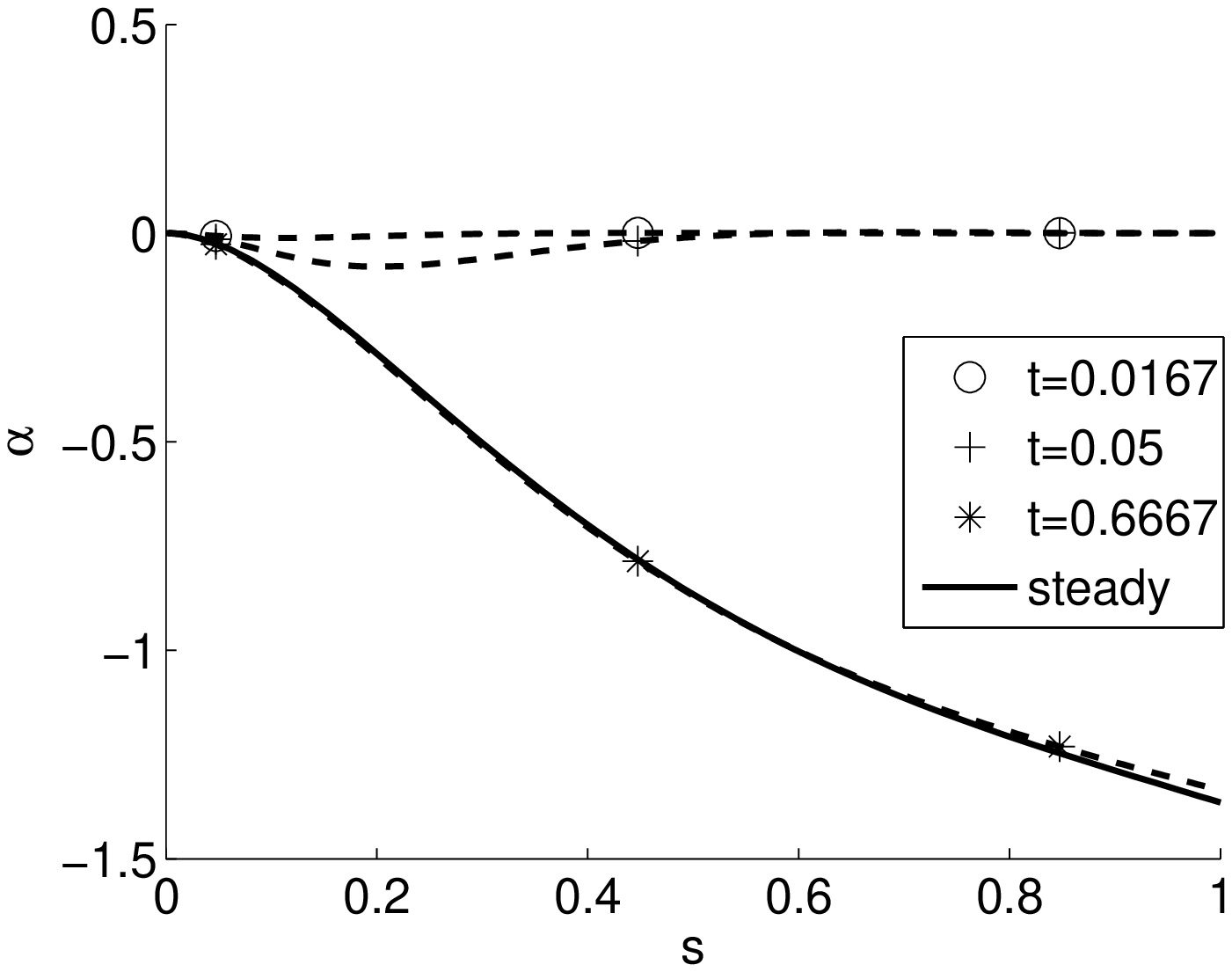}
\includegraphics[scale=0.425]{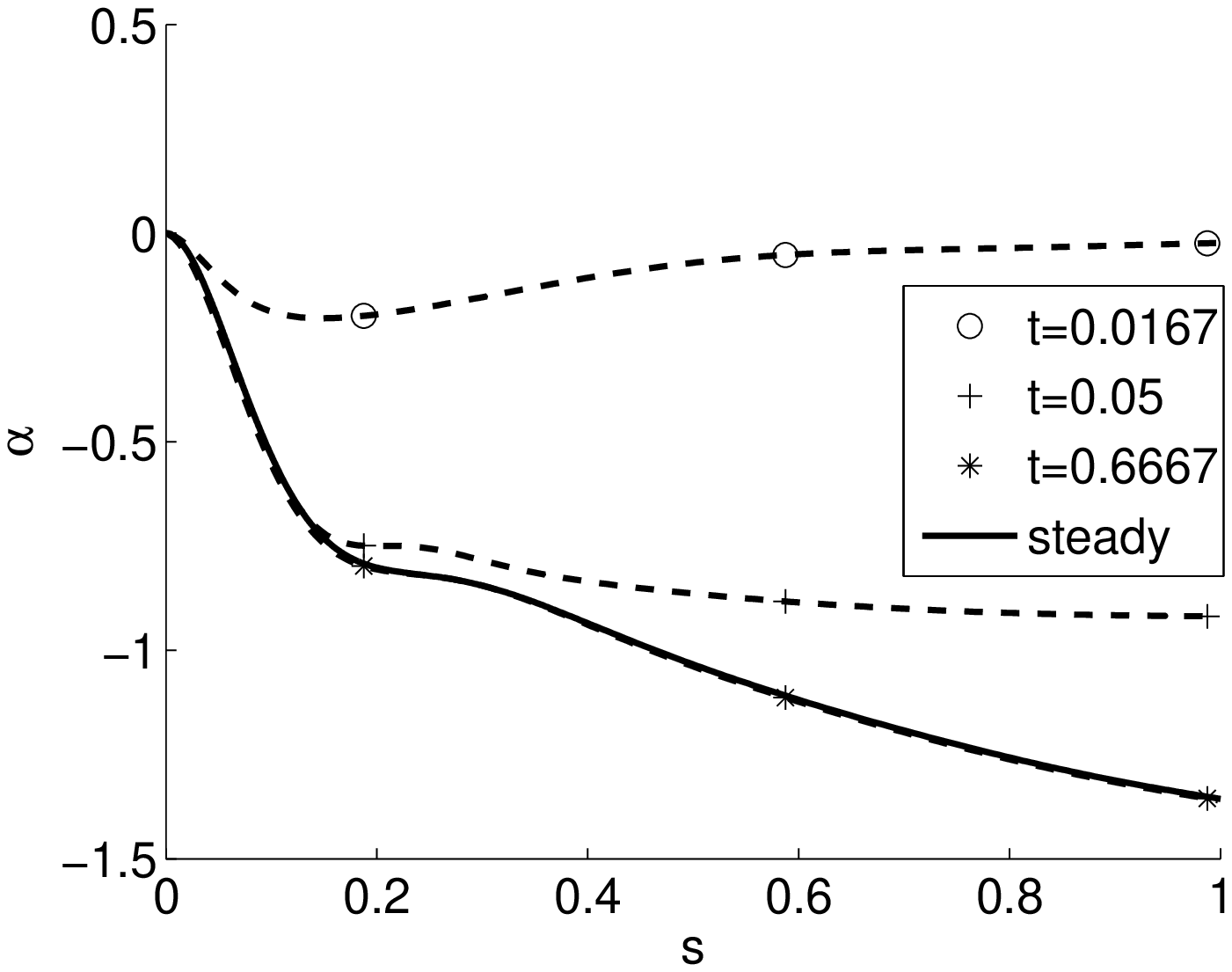}\\
\includegraphics[scale=0.425]{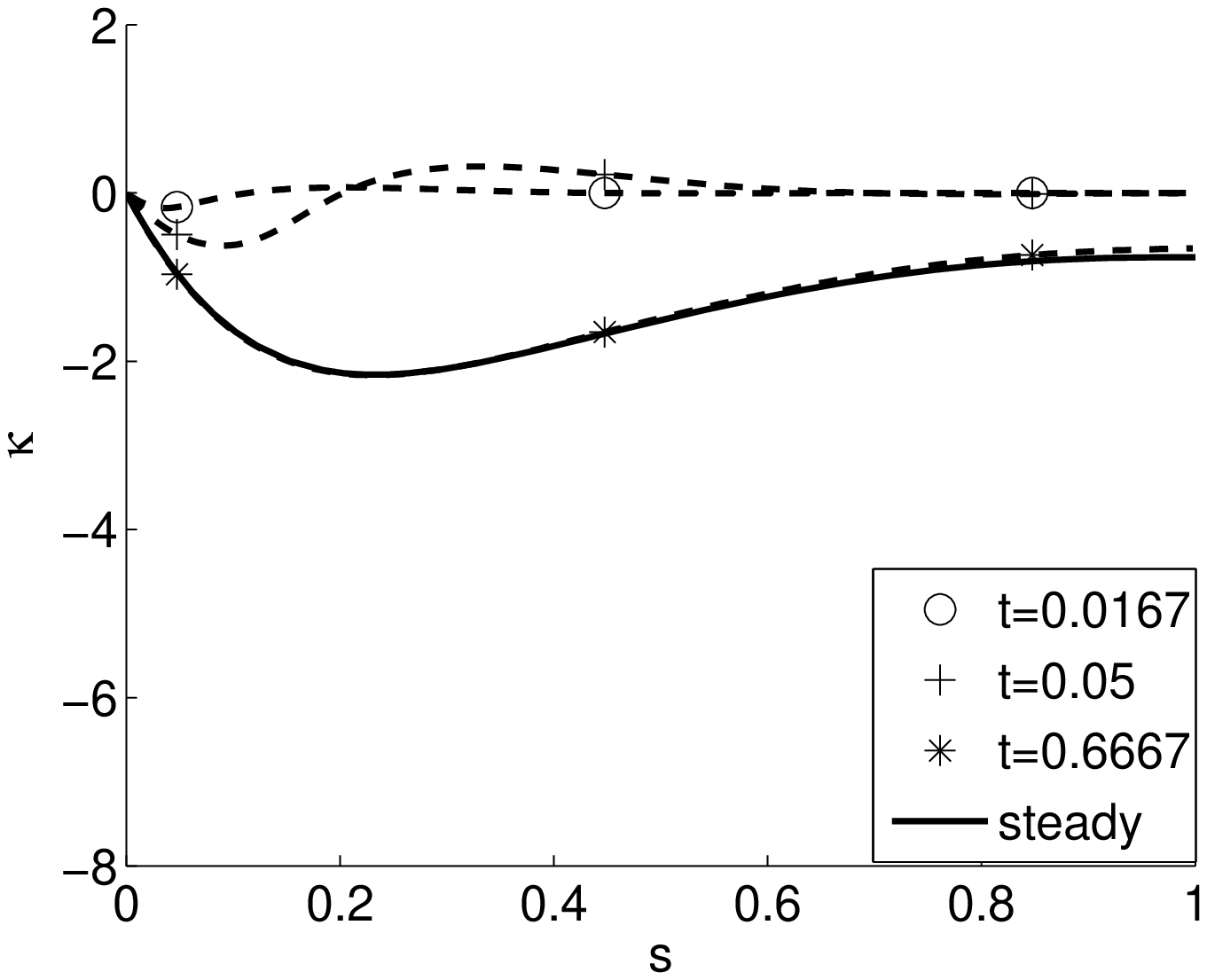}
\includegraphics[scale=0.425]{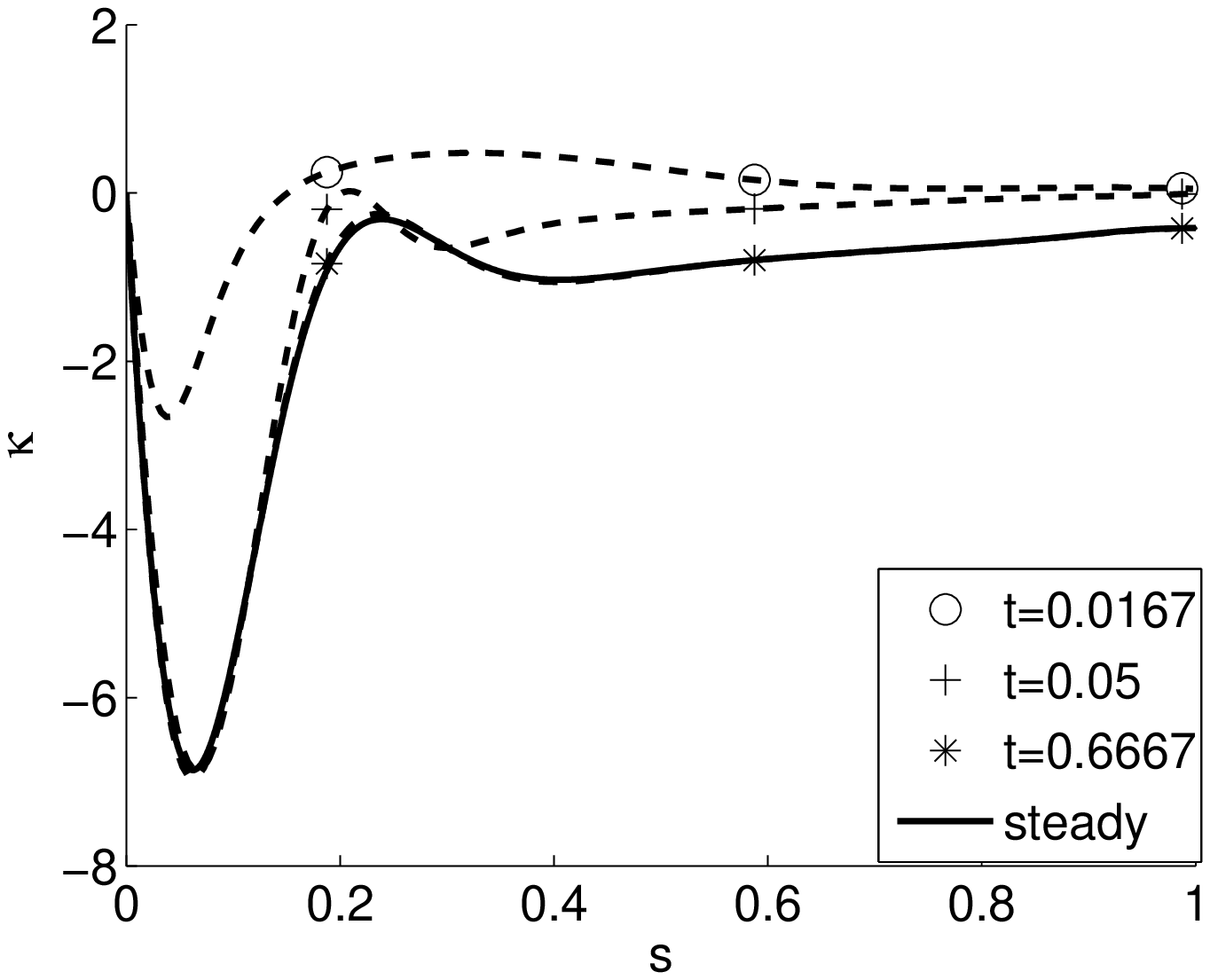}\\
\includegraphics[scale=0.425]{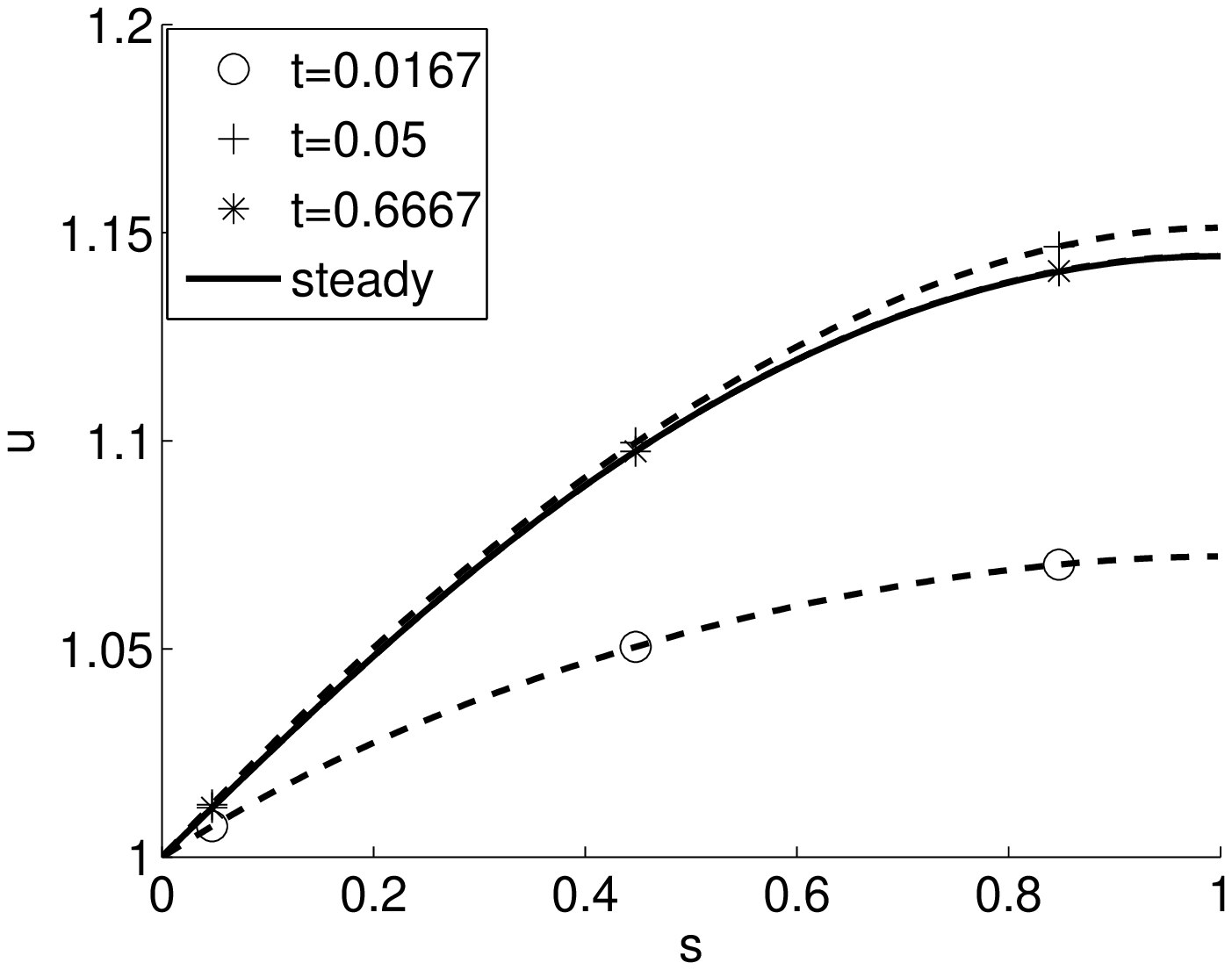}
\includegraphics[scale=0.425]{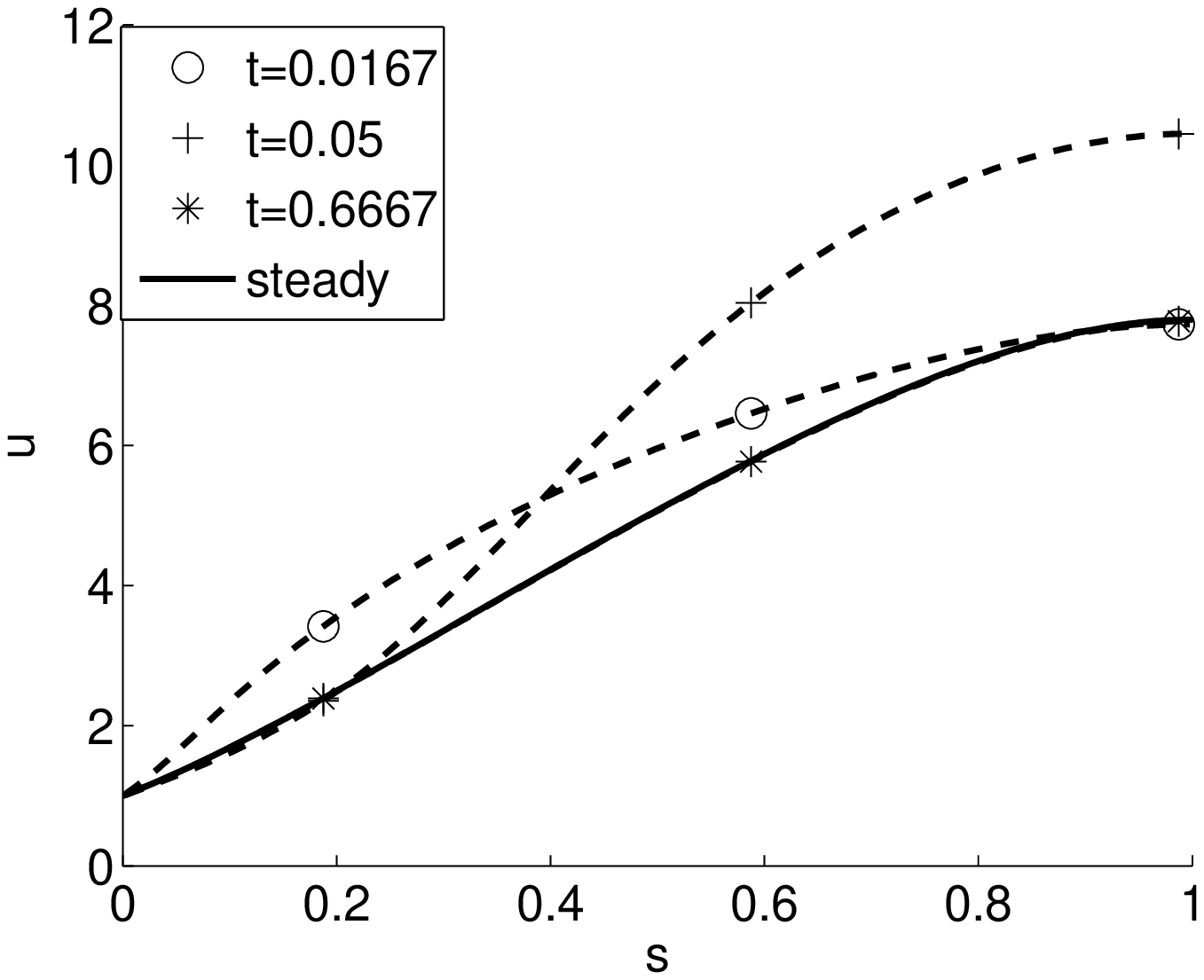}
\caption{\label{fig:euler} Temporal evolution of rod (three depicted times) in comparison to steady result of \cite{arne:p:2011} for inflow-outflow set-up in 2d. \textit{Top to bottom:} $\mathsf{r}$, $\alpha$, $\kappa$, $u$. Jet parameters are $\ell=1$, $\epsilon=0.1$, $\mathrm{Re}=1$. \textit{Left:} $\mathrm{Rb}=1$ (string is applicable). \textit{Right}: $\mathrm{Rb}=0.1$ (string fails). Note that for $u$ the scaling of the axes differs.}
\end{figure}

In this section we demonstrate the applicability of the rod model for simulating spinning processes and investigate the relevance of the instationary effects. For this purpose we present numerical results for two-dimensional and three-dimensional rotational spinning. The two-dimensional rotational spinning under the neglect of gravity can be considered as a bench-mark test scenario that has been often used in literature -- e.g.\ for the study of strings in \cite{wallwork:p:2002,goetz:p:2008,panda:d:2006,panda:p:2008,hlod:d:2009} and of stationary rods in \cite{arne:p:2010,arne:p:2011}. In this scenario ($\mathrm{Fr}\rightarrow \infty$) the rotation matrix $\mathsf{R}$ can be parameterized in terms of a single angle $\alpha\in[-\pi/2,0]$, see Fig.~\ref{fig:1}. Moreover, the number of variables reduce while the initial-boundary value problems keep their characteristic structure. The respective model equations for the Set-ups A and B in 2d are stated for completeness in the Appendix, Eqs.~\eqref{eq:L_2d} and \eqref{eq:E_2d}.

We start with the study of the longtime behavior of the instationary rods in comparison to the known stationary results of \cite{arne:p:2010,arne:p:2011}. Therefore, we consider Set-up B, inflow-outflow with fixed domain in Eulerian parameterization in 2d. The length ratios are exemplary chosen as $\ell=1$ and $\epsilon=0.1$. For the instationary rod we use the straight jet as initialization at $t=0$. The temporal evolution of the rod quantities are illustrated by help of three depicted time points in Fig.~\ref{fig:euler} showing two different sets of parameters. We clearly observe the convergence of the instationary solutions to the stationary ones as time increases, $t\rightarrow \infty$. The case $\mathrm{Re}=1=\mathrm{Rb}$ (Fig.~\ref{fig:euler}, left) lies in the parameter regime where also the string model is applicable. As $\epsilon\rightarrow 0$ the rod solution coincides with the string solution in consistency to the theoretical results (rod-to-string-convergence proof in \cite{arne:p:2011}). The other case $\mathrm{Re}=1$, $\mathrm{Rb}=0.1$ represents a jet of same viscosity, but exposed to faster rotations. In this regime the strings fail. Figure~\ref{fig:euler} shows here instationary jet simulations which can be similarly performed for the general three-dimensional inflow-outflow set-up with $\mathrm{Fr}<\infty$.

\begin{figure}[b]
\includegraphics[scale=0.425]{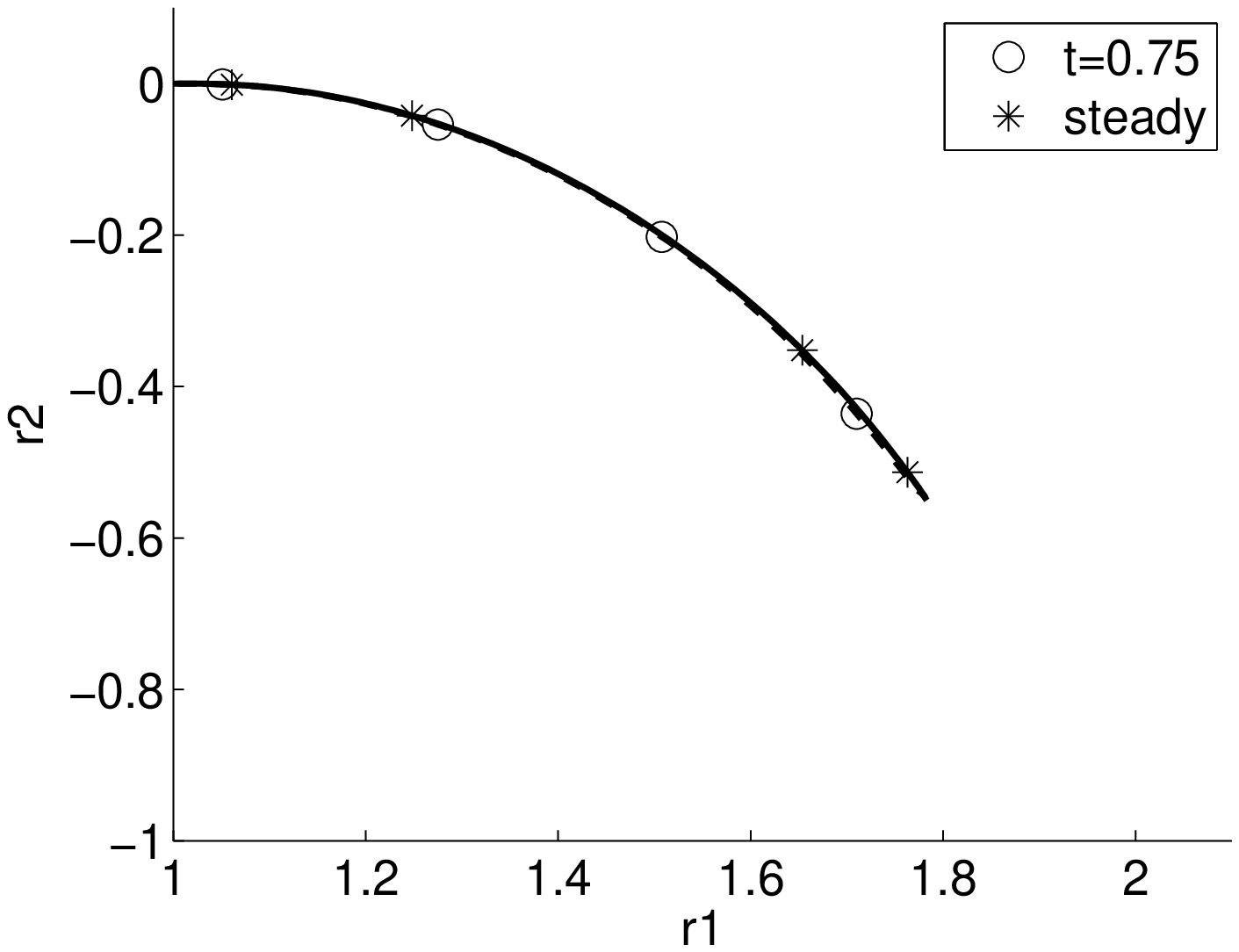}
\includegraphics[scale=0.425]{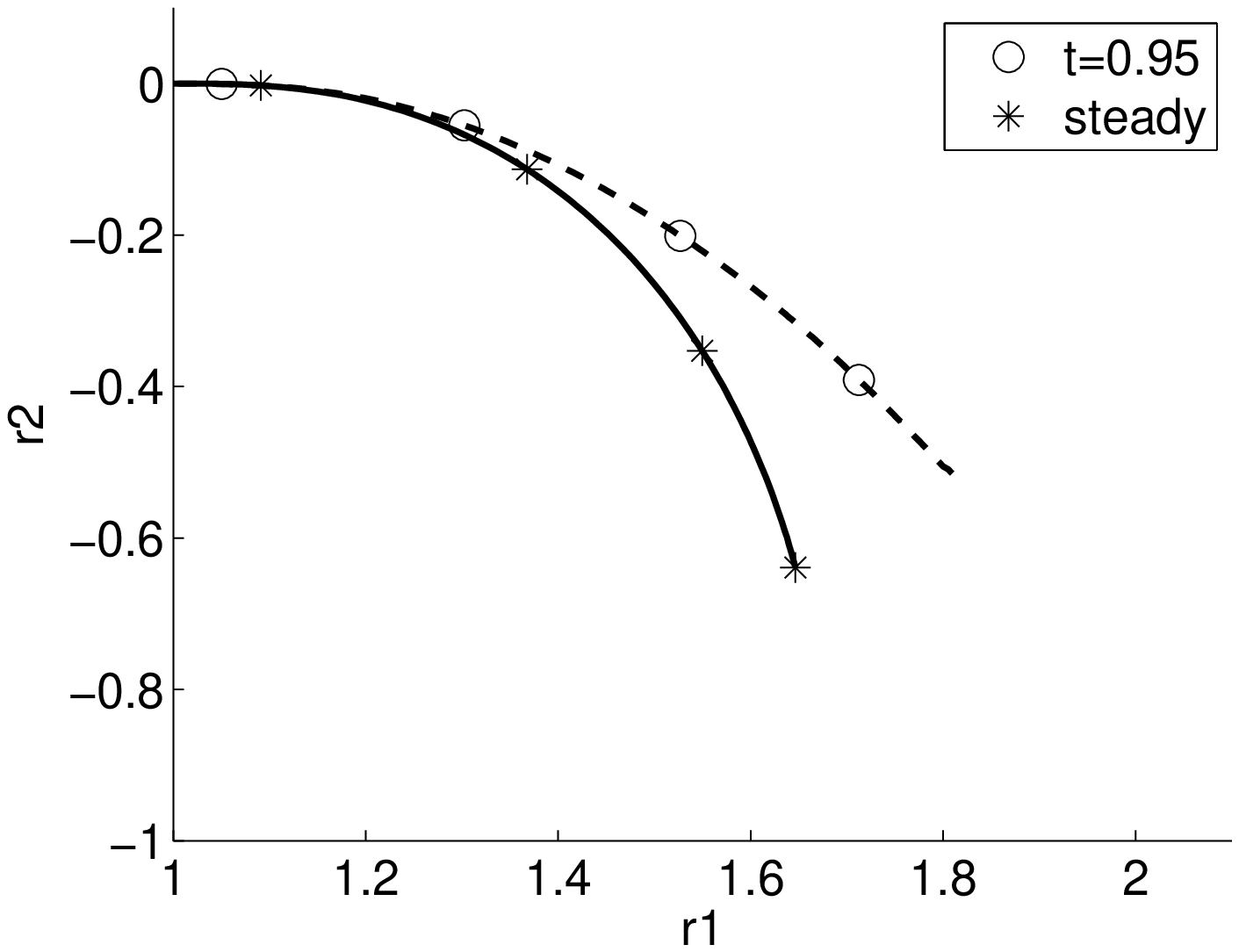}
\caption{\label{fig:vanish_vis} Time-dependencies of center-line for viscous jet spinning up to length $\ell$ (Set-up A) in comparison to stationary result computed with same fixed length $\ell$ in Set-up B. Jet parameters in 2d are $\epsilon=0.1$, $\mathrm{Rb}=1$. \textit{Left:} $\mathrm{Re}=100$ (jet growing along stationary curve). \textit{Right:} $\mathrm{Re}=1$ (dynamic curve approaching stationary behavior at nozzle for $t$ large).}
\end{figure}

Time-dependencies play a crucial role for highly viscous jets. Considering the spinning of highly viscous jets (Set-up A, inflow with free jet end) the instationary jet center-line is an important feature as experiments \cite{wong:p:2004} and corresponding instationary string simulations \cite{decent:p:2009,marheineke:p:2009,panda:p:2008} show. Figure~\ref{fig:vanish_vis} illustrates the well-known effect for the rod model. Whereas for inviscid flows (large $\mathrm{Re}$) the jet grows along a trajectory that coincides with the stationary jet curve computed for a certain length (according to Set-up B), the center-line for a viscous flow (small and moderate $\mathrm{Re}$) is clearly dynamic. However, for long-time it approaches to the stationary jet behavior near the nozzle. So, the stationary simulations of the inflow-outflow set-up turn out to be very suitable for studying the jet's properties close to the spinning nozzle. This hypothesis was already used for the design of production processes of technical textiles in \cite{arne:p:2012}, and it is now confirmed for all parameter regimes. 

\begin{figure}[t]
\includegraphics[scale=0.425]{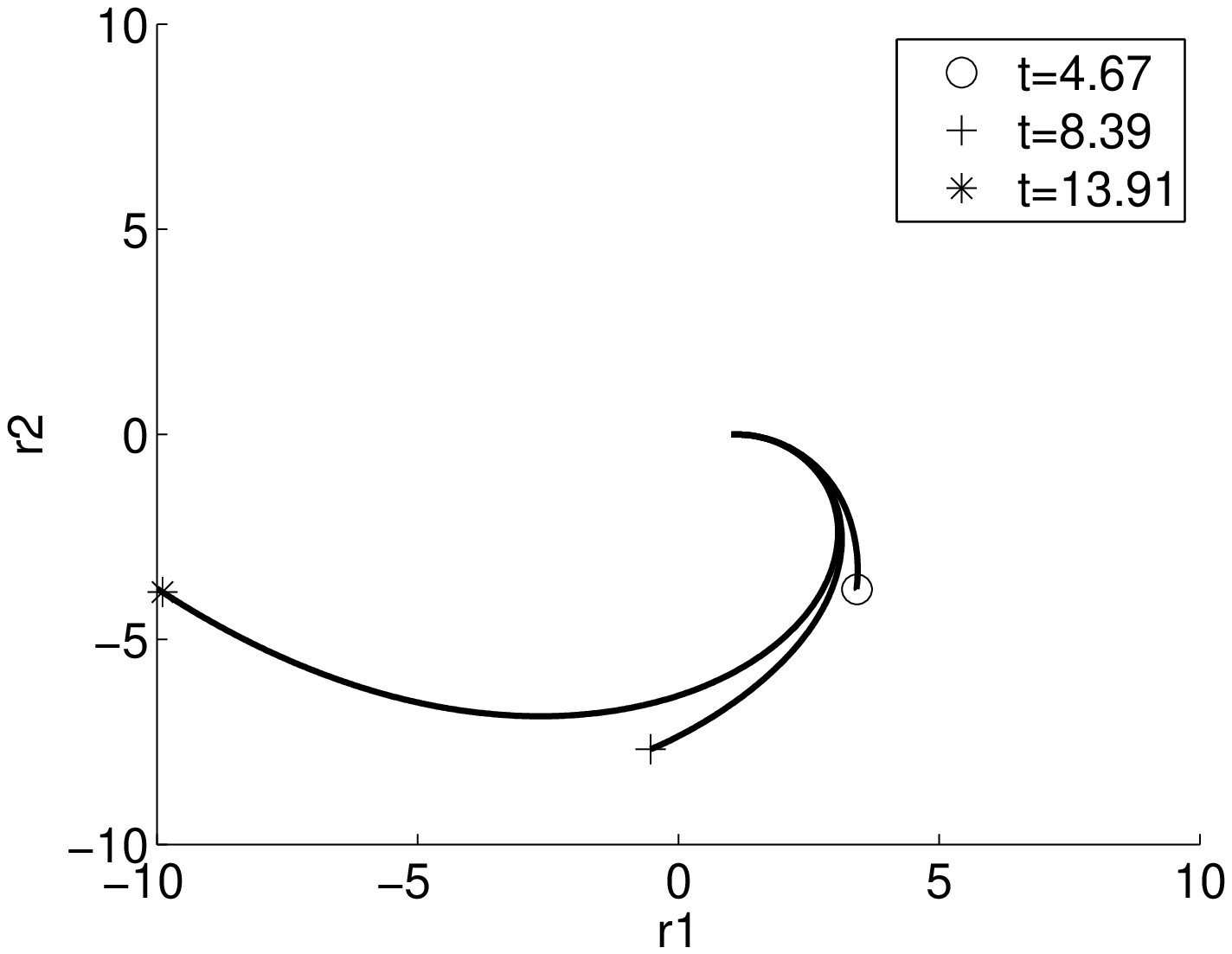}
\includegraphics[scale=0.425]{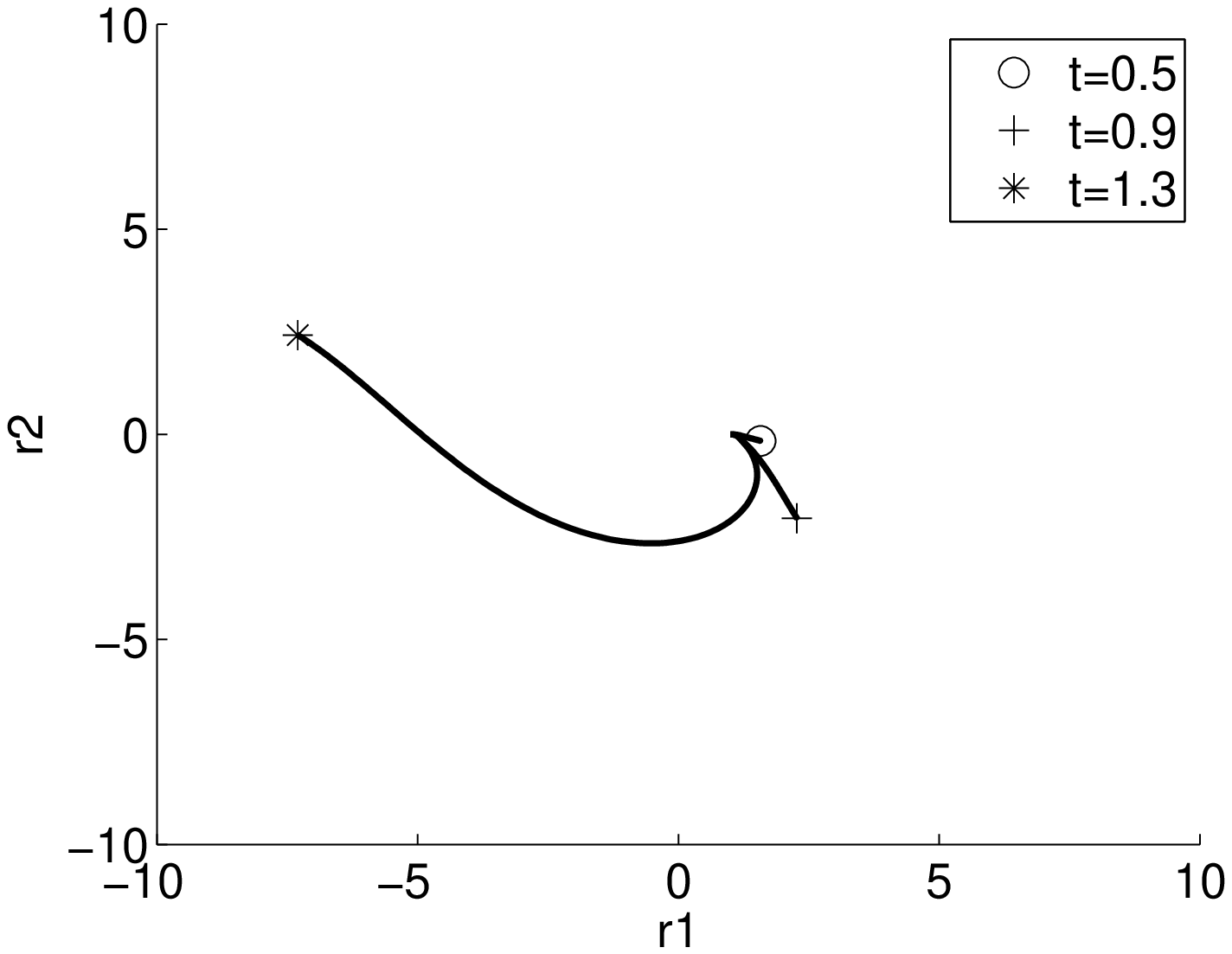}
\caption{\label{fig:vergl_panda_2d} Growing jet with free end (Set-up A in 2d), $\epsilon=0.1$. \textit{Left:} $\mathrm{Re}=1$, $\mathrm{Rb}=4$ (classical string regime, cf.\ string results \cite{panda:d:2006}). \textit{Right:} $\mathrm{Re}=\mathrm{Rb}=0.1$. Note that depicted time points are chosen with respect to the dynamics.}
\end{figure}

\begin{figure}[b]
\hspace*{0.5cm}\includegraphics[scale=0.475]{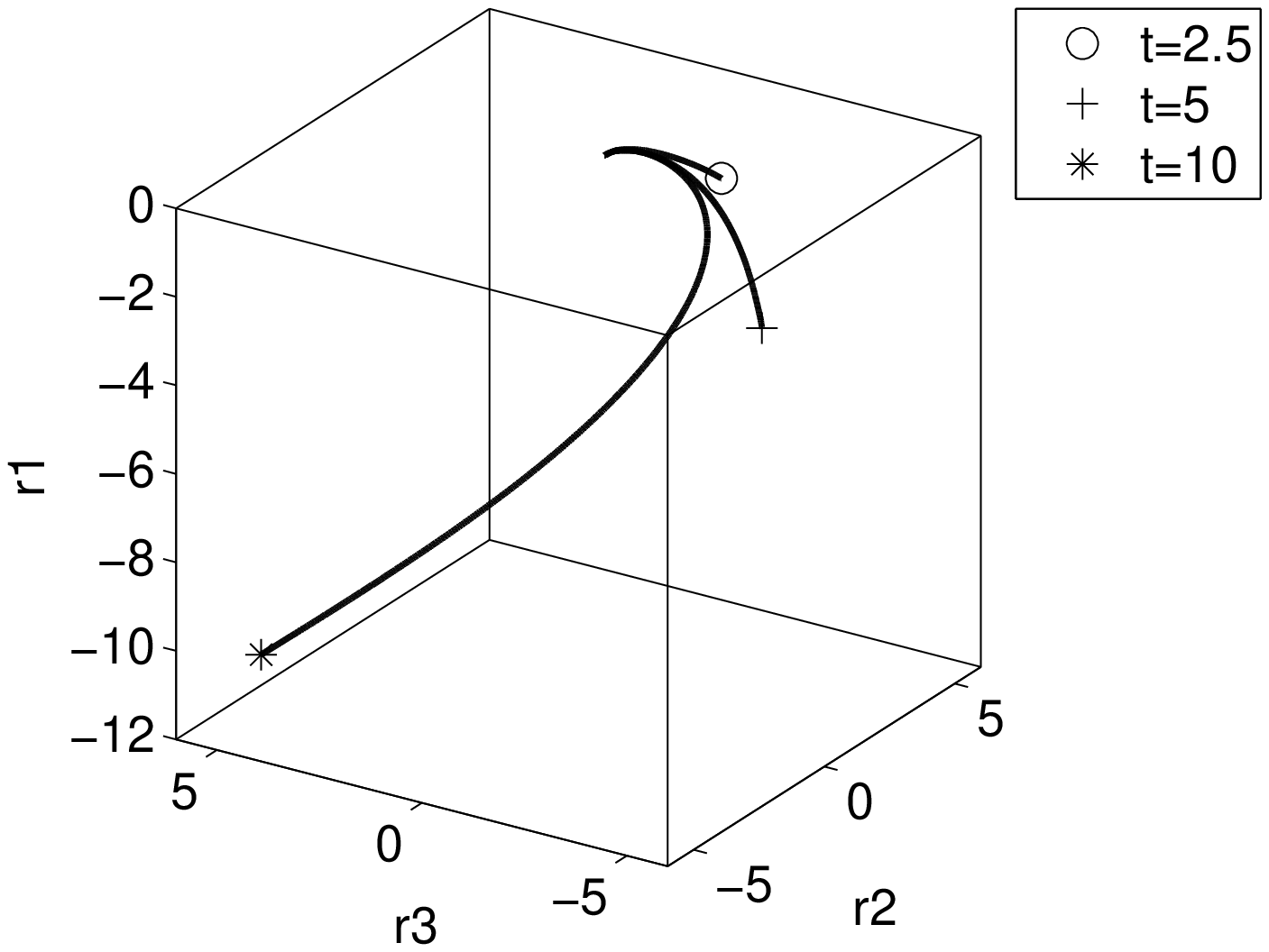}
\hspace*{-0.5cm}\includegraphics[scale=0.475]{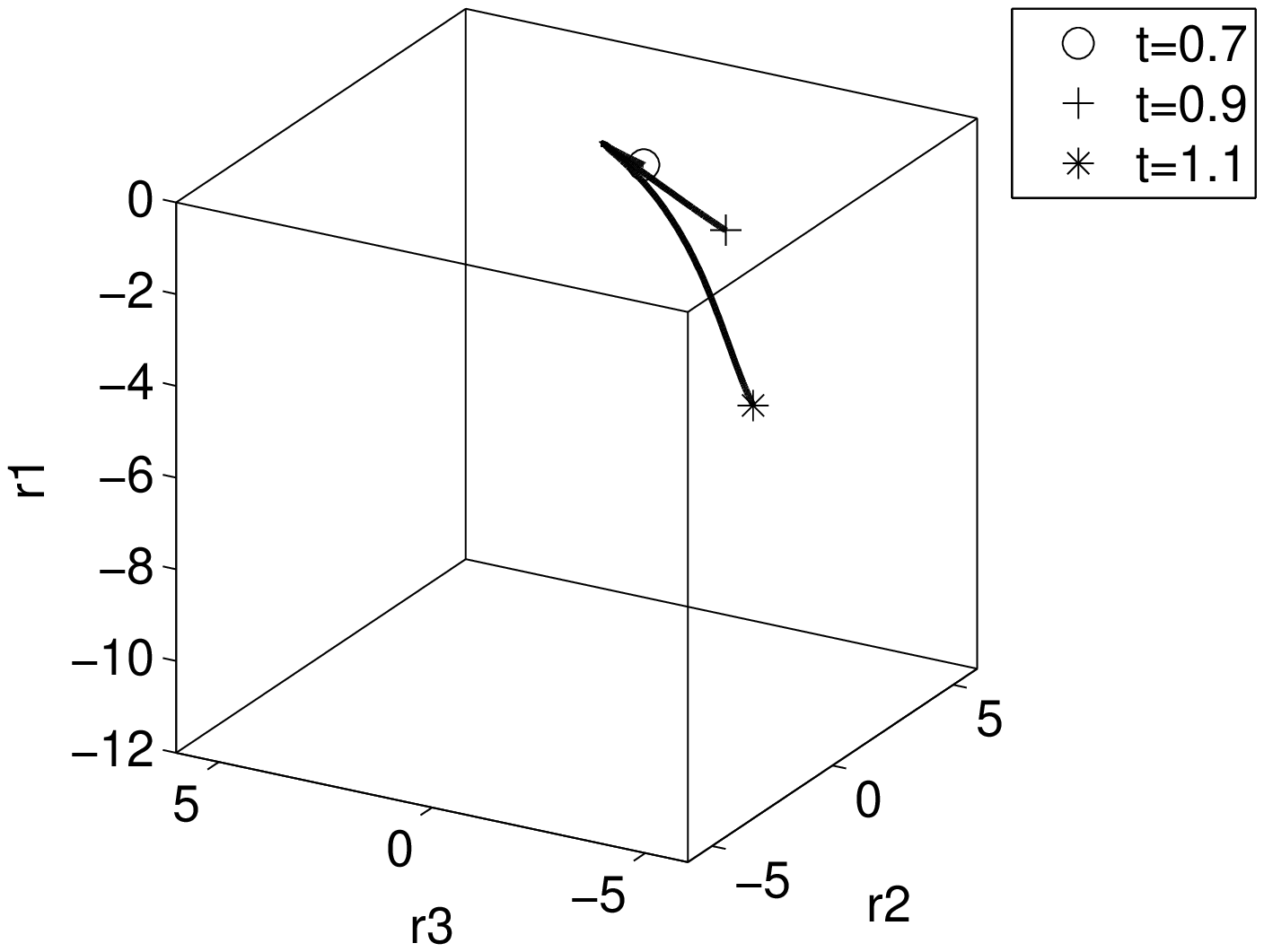}\\
\includegraphics[scale=0.425]{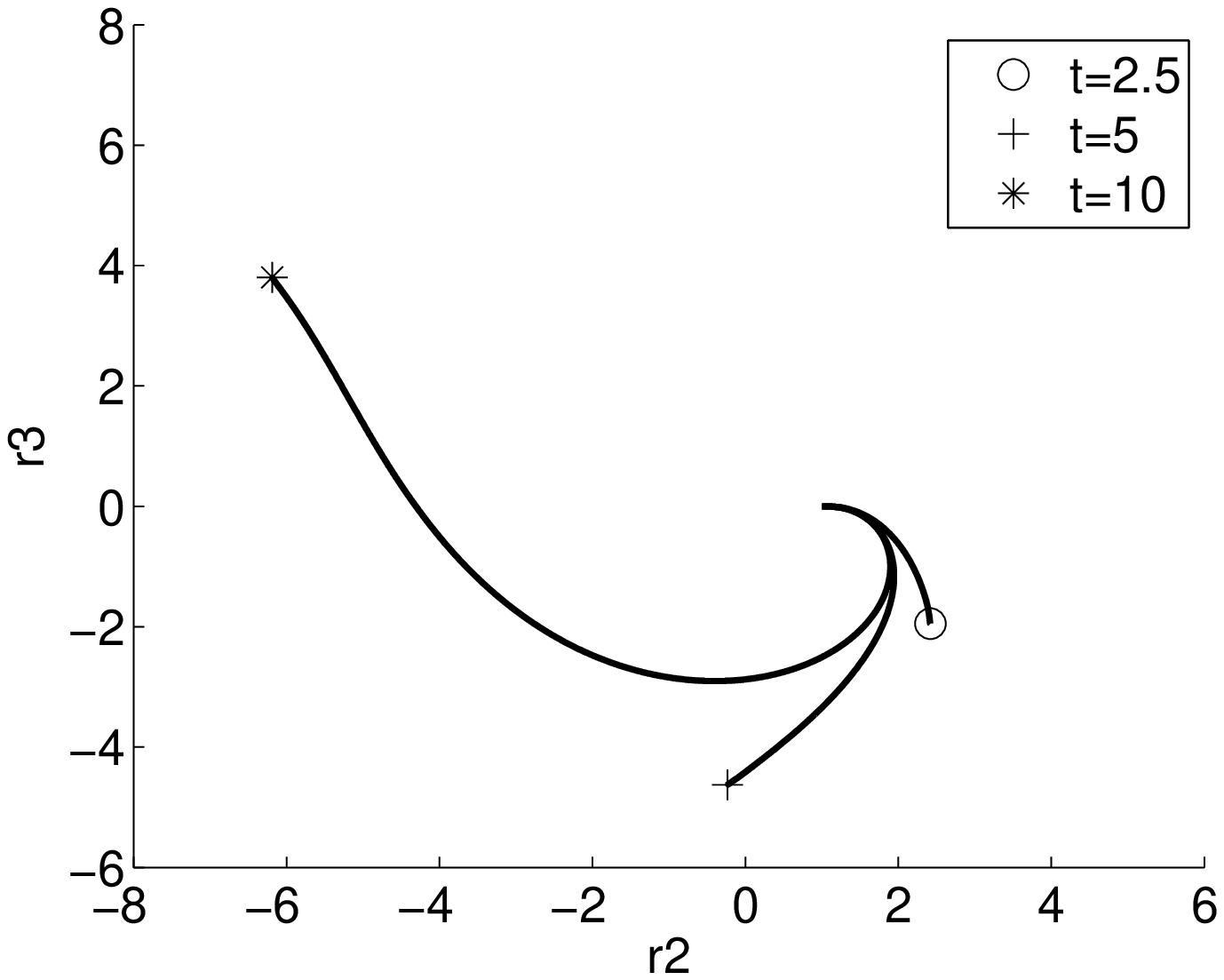}
\includegraphics[scale=0.425]{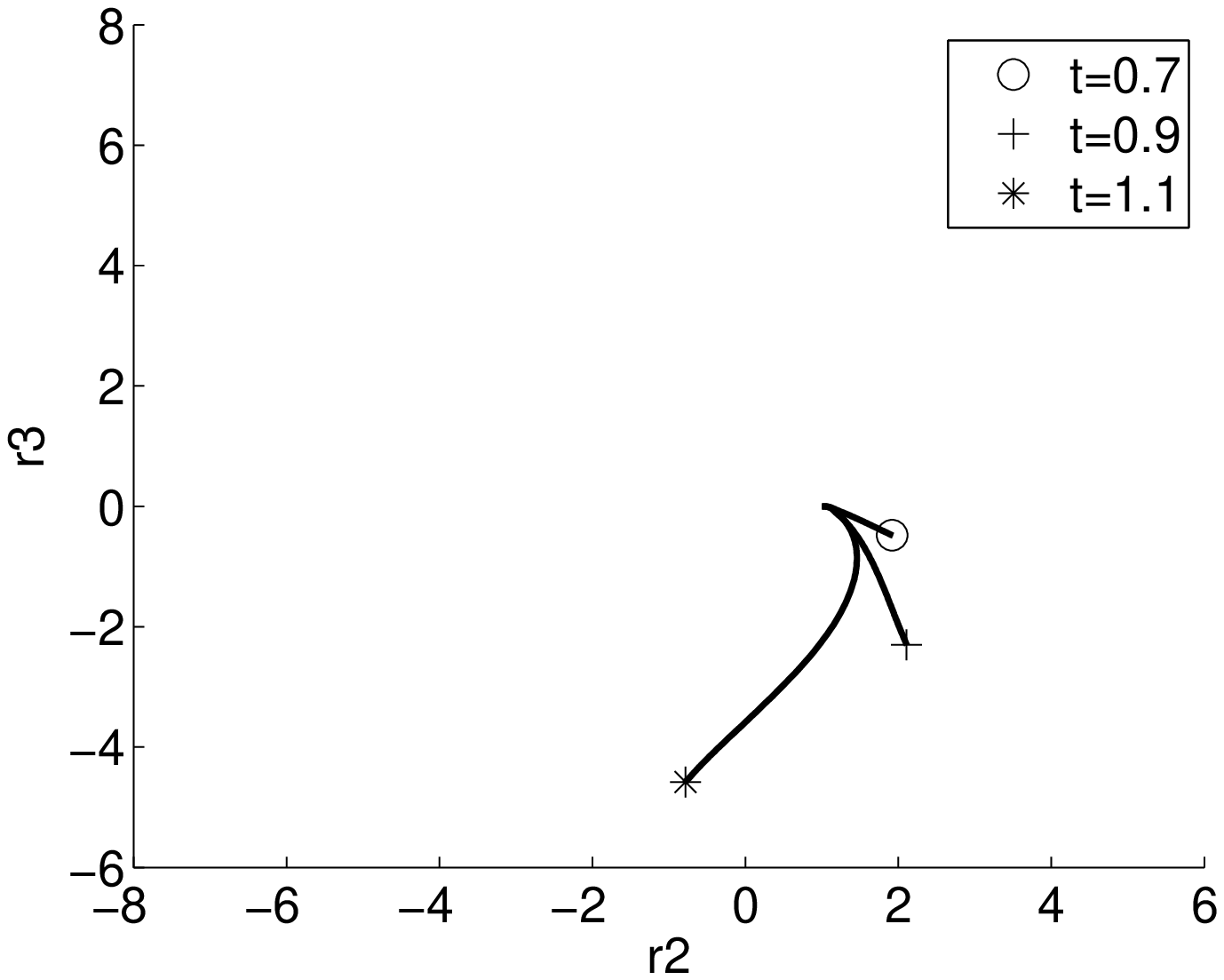}
\caption{\label{fig:vergl_panda} Growing jet with free end, Set-up A in 3d with 2d projections of top view \textit{(bottom)}, $\epsilon=0.1$. \textit{Left:} $\mathrm{Re}=1$, $\mathrm{Rb}=2$, $\mathrm{Fr}=2$ (classical string regime, cf.\ string results \cite{panda:p:2008}). \textit{Right:} $\mathrm{Re}=\mathrm{Rb}=\mathrm{Fr}=0.1$. Note that depicted time points are chosen with respect to the dynamics.}
\end{figure}

We come now to the results for the jet spinning (Set-up A) with arbitrarily chosen parameters and free end, whose efficient numerical handling is of main importance and interest for industrial applications in future. Figure~\ref{fig:vergl_panda_2d} presents a growing jet in 2d for two different sets of parameters. The case $\mathrm{Re}=1$, $\mathrm{Rb}=4$ (Fig.~\ref{fig:vergl_panda_2d}, left) has been already tackled by Panda \cite{panda:d:2006} using the string model. We use here the same depicted time points for the rod with $\epsilon=0.1$, the results are in very good agreement as $\epsilon \rightarrow 0$. The other case $\mathrm{Re}=0.1=\mathrm{Rb}$ (Fig.~\ref{fig:vergl_panda_2d}, right) lies outside of the applicability regime of the strings. So far, no respective simulations did exist. As expected the instationary Cosserat rod model opens the full parameter range to the simulation. This also holds true for the three-dimensional rotational spinning including gravity. Figure~\ref{fig:vergl_panda} shows the jet dynamics with respect to two arbitrary parameter tuples: one is chosen in the (classical) string regime, the other one outside of it. The parameter tuple associated to the string regime is taken from \cite{panda:p:2008}, the results of rod and string model agree as in 2d. The string model in general fails when facing high forces exposed to (highly) viscous jets (small $\mathrm{Re}$, $\mathrm{Rb}$, $\mathrm{Fr}$). The large deformations imply boundary layers which cause singularities in the solution and a break-down of theory (model) and numerics. The rod model overcomes this limitation since it is an $\epsilon$-regularization of the string. This fact is strict in theory for $\epsilon \neq 0$ and holds numerically as long as $\epsilon$ is evidently larger than zero (machine precision). As for the performance of the simulations, we already mentioned in Remark~\ref{rem:grid} that the applied discretization ($\triangle t$, $\triangle \sigma$) depends crucially on the considered problem parameters. Smaller parameters imply faster, larger changes in the dynamics and higher elongation $e$ which require a finer resolution. Theoretically, all parameter settings could be computed by help of our proposed scheme, but the simulations are practically restricted to problems with moderate elongation $e\leq 50$ due to the drastically increasing computational effort. This is no drawback for many spinning processes. But for example in industrial processes (like melt-blown) that are driven by turbulent air flows much higher elongations are observed.

%%%%%%%%%%%%%%%%%%%%%%%%%%%%%%%%%%%%%%%%%%%%%%%%%%%%%%%%%%%%%%%%%%%%%%%%%%%%%%%%%%%%%%%

\section{Conclusion}

The simulation of viscous jet spinning requires the efficient numerical handling of a slender enlarging flow domain with free end that is (highly) dynamic due to acting forces. We have proposed a finite volume approach with Radau time integration for the viscous Cosserat rod model consisting of a system of partial and ordinary differential equations that becomes stiff for small slenderness ratio $\epsilon$. The Cosserat rod is an $\epsilon$-regularization of the well-established asymptotic string models whose applicability are restricted to certain parameter settings. We have demonstrated the rod's superiority and large potential in view of industrial applications by performing instationary simulations of rotational spinning under gravity for arbitrary parameter ranges, for which the hitherto investigations in literature longed and failed. The work has also addressed the open question of a free end.

The established numerical scheme allows the easy incorporation of further practically relevant effects, like temperature dependencies and aerodynamic forces. 
This will be proceeded in future. The time and space discretizations depend on the considered problem parameters: small parameters cause fast, large deformations and high elongation which require a fine resolution. Due to the drastically increasing computational effort the use of our scheme is practically limited to problems with moderate elongations, so far. In view of turbulence-driven processes that yield very high elongations we intend to get rid of this bottle-neck by investigating appropriate refinement strategies.

%%%%%%%%%%%%%%%%%%%%%%%%%%%%%%%%%%%%%%%%%%%%%%%%%%%%%%%%%%%%%%%%%%%%%%%%%%%%%%%%%%%%%%%

\section*{Appendix}
\renewcommand{\theequation}{A.\arabic{equation}}

For completeness we state here the model simplifications of \eqref{eq:L_inflow} and \eqref{eq:E} for the two-dimensional rotational spinning scenario where gravity is neglected ($\mathrm{Fr}\rightarrow \infty$), cf.\ Fig.~\ref{fig:1}. The rotation is parameterized in terms of a single angle $\alpha\in[-\pi/2,0]$.
Moreover, we set $\mathsf{v}=(v_2,v_3)$, $\mathsf{v}^\perp=(-v_3,v_2)$, $\omega=\omega_1$, and all other quantities analogously. With $\Omega=\Omega_1$ and rotation matrix
\begin{align*}
\mathsf{R}(\alpha)=\left(\begin{array}{c c} \sin \alpha & -\cos \alpha \\ \cos \alpha & \sin \alpha \end{array} \right)
\end{align*}
we obtain after renumbering, i.e.\ $(z_2,z_3)$ becomes $(z_1,z_2)$ for all $\mathsf{z}$, the following two-dimensional systems.

%%%%%%%
\subsubsection*{Set-up A in 2d: inflow in Lagrangian parameterization, cf.~\eqref{eq:L_inflow}}

\begin{align} \label{eq:L_2d}
\mathsf{R}(\alpha) \cdot \partial_t \mathsf{\breve{r}} &=
\mathsf{v}\\ \nonumber
\partial_t \alpha  &= \mathsf{\omega}\\ \nonumber
\partial_t (e\mathsf{e_2}) & = 
\partial_{\sigma} \mathsf{v} + \kappa \mathsf{v}^\perp + \mathsf{\omega}e \mathsf{e_1}\\ \nonumber 
\partial_t \mathsf{\kappa} & = 
\partial_{\sigma} \mathsf{\omega} \\ \nonumber
\partial_t\mathsf{v} &=\frac{1}{\mathrm{Re}}(
\partial_{\sigma} \mathsf{n} + \mathsf{\kappa}\mathsf{n}^\perp) -  \omega {\sf v}^\perp-\frac{2}{\mathrm{Rb}} {\sf v}^\perp
+ \frac{1}{\mathrm{Rb^2}} {\sf R}(\alpha)\cdot {\sf \breve r}\\ \nonumber
\partial_t \frac{\mathsf{\omega}}{e} &= 
\frac{4}{\mathrm{Re}}\partial_\sigma m -\frac{16}{\epsilon^2 \mathrm{Re}}{n_1} + \frac{1}{\mathrm{Rb}}\frac{\partial_te}{e^2}
\end{align}
with
\begin{align*}
n_2=3\frac{1}{e^2}(\partial_\sigma v_2+\kappa v_1), \quad \quad \quad m=\frac{3}{4}\frac{\partial_\sigma \omega}{e^3}\,.
\end{align*}

%%%%%%
\subsubsection*{Set-up B in 2d: inflow-outflow in Eulerian parameterization, cf.~\eqref{eq:E}}

\begin{align} \label{eq:E_2d}
\mathsf{R}(\alpha) \cdot \partial_t \mathsf{\breve{r}} &=
\mathsf{v} -u \mathsf{e_2}\\ \nonumber
\partial_t \alpha  &= \mathsf{\omega}-u\kappa \\ \nonumber
\partial_s (u\mathsf{e_2}) & = 
\partial_s \mathsf{v} + \kappa \mathsf{v}^\perp + \mathsf{\omega} \mathsf{e_1}\\ \nonumber 
\partial_t \mathsf{\kappa} + \partial_s (u \mathsf{\kappa}) & = 
\partial_s \mathsf{\omega} \\ \nonumber
\partial_t A + \partial_s (u A) & =  0\\ \nonumber
\partial_t(A \mathsf{v}) + \partial_s(u A \mathsf{v}) &=\frac{1}{\mathrm{Re}}(
\partial_s \mathsf{n} + \mathsf{\kappa}\mathsf{n}^\perp) -  A \omega {\sf v}^\perp
-\frac{2}{\mathrm{Rb}} A{\sf v}^\perp
+ \frac{1}{\mathrm{Rb^2}}A {\sf R}(\alpha)\cdot {\sf \breve r}\\ \nonumber
\partial_t (A^2\mathsf{\omega}) +  \partial_s(u A^2 {\omega}) &= 
\frac{4}{\mathrm{Re}}\partial_s m -\frac{16}{\epsilon^2 \mathrm{Re}}{n_1} + \frac{1}{\mathrm{Rb}}A^2\partial_s u
\end{align}
with
\begin{align*}
n_2=3A\partial_s u, \quad \quad m=\frac{3}{4}A^2  \partial_s \mathsf{\omega}.
\end{align*}
For the simulations in Section~\ref{sec:4}, the system \eqref{eq:E_2d} is initialized with the straight jet that leaves the nozzle perpendicularly,
\begin{align*}
r_1(s,0)&=s+1 && r_2(s,0)=0 && \alpha(s,0)=0 \\
\kappa(s,0)&=0 && u(s,0)=1 && n_1(s,0)=0 \\
A(s,0)&=1 && A\mathsf{v}(s,0)=(0,1) && A^2\omega(s,0)=0\,.
\end{align*}

%%%%%%%%%%%%%%%%%%%%%%%%%%%%%%%%%%%%%%%%%%%%%%%%%%%%%%%%%%%%%%%%%%%%%%%%%%%%%%%%%%%%%%%%
\quad\\
{\sc Acknowledgments.} This work has been supported by German Bundesministerium f\"ur Bildung und Forschung (BMBF), Schwerpunkt "Mathematik f\"ur Innovationen in Industrie und Dienstleistungen", Projekt 03MS606. 

%%%%%%%%%%%%%%%%

\end{document}